\documentclass{article}
\usepackage{amsmath}
\usepackage{graphicx}
\usepackage{amsfonts}
\usepackage{amssymb}

\begin{document}

\title{Anharmonic oscillator, negative dimensions and inverse factorial convergence
of large orders to the asymptotic form}
\author{P.V. Pobylitsa\\\emph{Petersburg Nuclear Physics Institute,}\\\emph{Gatchina, St.~Petersburg, 188300, Russia}}
\date{}
\maketitle

\begin{abstract}
The spectral problem for $O(D)$ symmetric polynomial potentials allows for a
partial algebraic solution after analytical continuation to negative even
dimensions $D$. This fact is closely related to the disappearance of the
factorial growth of large orders of the perturbation theory at negative even
$D$. As a consequence, certain quantities constructed from the perturbative
coefficients exhibit fast inverse factorial convergence to the asymptotic
values in the limit of large orders. This quantum mechanical construction can
be generalized to the case of quantum field theory.

\end{abstract}

\section{Introduction}

\setcounter{equation}{0} 

The $O(D)$ symmetric anharmonic oscillator
\begin{equation}
H=\frac{1}{2}\sum\limits_{i=1}^{D}\left(  p_{i}^{2}+x_{i}^{2}\right)
+V\left(  \sum\limits_{i=1}^{D}x_{i}^{2}\right)  \label{H-general}%
\end{equation}
with a polynomial potential $V(r^{2})$ is a good theoretical toy for testing
various ideas used in quantum field theory. In particular, one should mention:

1) large orders of perturbation theory
\cite{BW-69,BW-73,Lipatov-77b,BLZ-77,SZ-79,DP-79,Zinn-Justin-81a,Zinn-Justin-81b,ZJ-04}%
,

2) fermion representations for boson theories in negative dimensions
\cite{DH-88,Dunne-89},

3) hidden symmetries, quasi-exactly solvable models
\cite{Turbiner-94,Shifman-ITEP,Ushveridze}, correspondence between integrable
quantum field theoretical models and ordinary differential equations
\cite{DT-99,BLZ-98,BLZ-03,DDT-07,DDGT-07}, correspondence between $O(D)$
symmetric and $D=1$ anharmonic oscillator systems
\cite{SZ-79,Andrianov-82,DPM-84,BG-93,Andrianov-07}.

4) Regge theory of complex angular momenta in nonrelativistic quantum
mechanics \cite{DeAlfaro-Regge,Collins-book,Newton-book}.

There is a vast literature where the anharmonic oscillator is studied from
various points of view. The aim of this paper is to bring some of these ideas
together. The paper consists of two parts. The first part is devoted to the
spectrum of the anharmonic oscillator at even negative $D$
\begin{equation}
D=0,-2,-4,-6,\ldots\label{D-even-negative}%
\end{equation}
understood in the sense of analytical continuation. A remarkable feature of
these values of $D$ is that a part of the spectrum can be found by solving
algebraic equations. In principle, this fact is known since long ago
\cite{DH-88} but as usual with exactly solvable models, one can arrive at the
same results using many different ways, and the alternative derivations
discussed here may deserve attention.

The second part of the paper deals with the asymptotic behavior of large
orders of the perturbation theory. As is well known, the perturbative series
for the anharmonic oscillator is divergent because of the factorial growth of
the coefficients. However, one can show that in even negative dimensions
(\ref{D-even-negative}) the factorial growth disappears and the perturbative
series becomes convergent. Taken alone, this fact is quite natural and not too
exciting. But this observation gives an interesting solution to another old problem.

Let us return to the case of general $D$ where we have the factorial growth of
perturbative coefficients. It is well known that the asymptotic behavior of
large-$k$ orders of the perturbation theory is reached very slowly because of
the $O(k^{-1})$ corrections that are usually rather large in practically
interesting cases. Therefore it would be rather interesting to construct
quantities for which the asymptotic regime is reached faster than $O(k^{-1})$.
This problem is especially acute in quantum field theoretical models where
only several first orders of the perturbation theory can be usually computed.
In this paper we construct and study special quantities whose approach to the
asymptotic regime is \emph{factorially fast} instead of the traditional slow
$O(k^{-1})$ convergence to the asymptotic form. It is interesting that the
same construction can be used in quantum field theory. In fact, this paper
appeared as an attempt to understand some superfast convergence effects that
are seen in quantum field theory \cite{PP-08}.

Before starting with a detailed derivation of the exact solution for the
spectrum of the anharmonic oscillator at negative even dimensions $D$, it
makes sense to describe the final result. For simplicity let us concentrate on
levels with zero angular momentum $l=0$.

At negative even $D$ a part of the discrete spectrum of energy $E$ is given by
roots of the algebraic equation
\begin{gather}
R_{M}(E)=0\,,\label{R-M-g-spectrum-eq}\\
M=-\frac{D}{2}=0,1,2,3,\ldots
\end{gather}
where $R_{M}(E)$ are certain polynomials of $E$ depending on potential $V$. In
this paper a rather elegant representation for these polynomials will be
derived
\begin{equation}
R_{M}(E)=\det\left[  J_{+}+V\left(  2J_{-}\right)  -E\right]  \,.
\label{R-N-det}%
\end{equation}
Here
\begin{equation}
J_{\pm}=J_{1}\pm iJ_{2}%
\end{equation}
are standard spin matrices for spin
\begin{equation}
j=\frac{M}{2}=-\frac{D}{4}=0,\frac{1}{2},1,\frac{3}{2},\ldots
\end{equation}

Several comments should be made about the history of the exact solution for
the anharmonic oscillator. The limit $D\rightarrow0$ was considered by Dolgov
and Popov \cite{DP-79}. From the point of view of representation
(\ref{R-N-det}) this case corresponds to spin $j=0$ with the polynomial
$R_{0}=-\mathrm{\,}E$ leading according to eq. (\ref{R-M-g-spectrum-eq}) to
only level controlled by the exact solution $E=0$. Although this case is
trivial from the spectral point of view, the work of Dolgov and Popov
\cite{DP-79} contains several interesting results. In particular, the wave
function corresponding to the level $E=0$ was computed in Ref. \cite{DP-79}.

The general case of arbitrary negative even $D$ was studied by Dunne and
Halliday in Ref. \cite{DH-88} where a method was suggested for a calculation
of polynomials describing the spectrum. In principle, this solves the problem
of the spectrum. However, some issues were not clarified. Testing some special
cases, the authors of \cite{DH-88} found an unexpected factorization of their
polynomials into ``elementary polynomials''. In the framework of the current
paper the ``elementary polynomials'' of Dunne and Halliday are nothing else
but polynomials $R_{M}(E,g)$ given by eq. (\ref{R-N-det}). An important role
in the disentanglement of the polynomial structure is played by a ``hidden''
$sl(2)$ symmetry which stands behind the spin representation (\ref{R-N-det})
and explains the miracles observed in Ref. \cite{DH-88}.

In this paper we discuss several derivations of polynomials (\ref{R-N-det}).
One method described in section \ref{sl-2-algebra-section} uses an explicit
expression for the Hamiltonian of the anharmonic oscillator in terms of
differential operators obeying $sl(2)$ algebra. Another way to polynomials
(\ref{R-N-det}) considered in Sec. \ref{Recusrion-relation-subsection} is
based on the analysis of a recursion relation for the coefficients of the
power series for the wave function. In the third method (Sec.
\ref{Analytical-PF-section}) one performs analytical continuation of the $l$
decomposition for the partition function. One more method (Sec.
\ref{Fermion-section}) uses a fermion representation for the partition
function continued analytically to negative dimensions. This approach is very
close to the method of original work \cite{DH-88} but we implement this method
so that the $sl(2)$ algebra remains explicit.

An important aspect of the problem of the anharmonic oscillator in negative
dimensions is the precise definition of negative dimensions. It is interesting
that the problem of analytical continuation to negative or complex dimensions
is equivalent to the problem of analytical continuation to complex angular
momenta which is a corner stone of Regge theory (Sec.
\ref{Regge-theory-subsection}).

As was already mentioned, at negative even $D$ we meet interesting phenomena
in large orders of the perturbation theory. The energy of the ground state of
the anharmonic oscillator has a perturbative expansion
\begin{equation}
E(g,D)=\sum\limits_{k=0}^{\infty}E^{(k)}(D)g^{k} \label{E-series-0}%
\end{equation}
with coefficients $E^{(k)}(D)$ which are polynomials of $D$ so that there are
no problems with analytical continuation of $E^{(k)}(D)$ to negative
dimensions. In the case of the quartic anharmonic oscillator%

\begin{equation}
V(r^{2})=\frac{1}{2}r^{2}+gr^{4} \label{H-quartic}%
\end{equation}
the large-$k$ asymptotic behavior of $E^{(k)}(D)$ is described by formula
\cite{BLZ-77}%
\begin{equation}
E^{(k)}(D)\overset{k\rightarrow\infty}{=}(-1)^{k+1}\Gamma\left(  k+\frac{D}%
{2}\right)  3^{k+\frac{D}{2}}\frac{2^{D/2}}{\pi\Gamma\left(  D/2\right)
}\left[  1+O(k^{-1})\right]  . \label{E-k-asymptotic-0}%
\end{equation}
An interesting feature of this formula is that the RHS vanishes at even
negative $D$ because of the factor $\left[  \Gamma\left(  D/2\right)  \right]
^{-1}$ on the RHS (\ref{E-k-asymptotic-0}). In section
\ref{No-factorials-subsection} we show that the factorial divergence of series
(\ref{E-series-0}) disappears at negative even values of $D$ so that the
perturbative series has a nonzero convergence radius.

Functions $E^{(k)}(D)$ are polynomials of $D$ of degree $k+1$ and they have
$k+1$ roots:%
\begin{equation}
E^{(k)}(\nu_{k,r})=0\quad(1\leq r\leq k+1).
\end{equation}
The last part of the paper is devoted to asymptotic properties of these roots
in the limit of large $k$. One could wonder why we should care about the roots
$\nu_{k,r}$. The answer is that these roots have a very interesting property:
in the limit $k\rightarrow\infty$ some subsets of the roots are convergent to
points $D=-4,-6,-8,\ldots$ and this convergence is factorially fast. In
quantum mechanics this superfast factorial convergence may be an amazing but
useless curiosity. However, in the context of quantum field theory analogous
effects are important. Indeed, in quantum field theory one is usually limited
to a rather small amount of perturbative terms. Four or five orders are
usually considered as a great achievement. Therefore the existence of
quantities with factorial convergence to the asymptotic form instead of the
slow $O(k^{-1})$ convergence of (\ref{E-k-asymptotic-0}) is extremely
interesting for field theoretical applications. The case of the anharmonic
oscillator allows us to study this phenomenon in detail.

\section{$O(D)$ symmetric Hamiltonians and $sl(2)$ algebra}

\setcounter{equation}{0} 

\label{sl-2-algebra-section}

\subsection{Schr\"{o}dinger equation}

\label{Schroedinger-section}

The $D$-dimensional Hamiltonian with a spherically symmetric potential
(\ref{H-general}) becomes in the polar coordinates%
\begin{equation}
H_{r}=\frac{1}{2}\left(  -\frac{d^{2}}{dr^{2}}-\frac{D-1}{r}\frac{d}%
{dr}\right)  +\frac{l(l+D-2)}{2r^{2}}+V(r^{2})\,.
\end{equation}
In $D\geq2$ dimensions, parameter $l$ runs over nonnegative integer values%
\begin{equation}
l=0,1,2,3,\ldots\quad(D\geq2)
\end{equation}

The degeneracy of the $l$ levels is given by%
\begin{equation}
m(D,l)=\frac{(2l+D-2)}{l!}\frac{\Gamma(D+l-2)}{\Gamma(D-1)}\,.
\label{l-degeneracy-positive}%
\end{equation}
The cases $D=1$, $D=2$ are somewhat exceptional. But they still can be
described by analytical continuation of (\ref{l-degeneracy-positive}) in $D$
(at fixed integer nonnegative $l$):

For $D=2$ and nonnegative integer $l$, we obtain%
\begin{equation}
m(2,l)=2-\delta_{l0}\,,
\end{equation}
which can interpreted in terms of the $\pm l$ degeneracy for $l\neq0$.

For $D=1$, we have%
\begin{equation}
m(1,l)=\delta_{l0}+\delta_{l1}%
\end{equation}
so that the only allowed values are $l=0,1$, and they correspond to the states
with positive and negative parity respectively.

After the separation of the factor $r^{l}$ from the wave function one arrives
at%
\begin{equation}
H_{\mathcal{D}}=r^{-l}H_{r}r^{l}=\frac{1}{2}\left(  -\frac{d^{2}}{dr^{2}%
}-\frac{\mathcal{D}-1}{r}\frac{d}{dr}\right)  +V(r^{2}) \label{H-D-def}%
\end{equation}
where%
\begin{equation}
\mathcal{D}=D+2l\,. \label{Delta-def}%
\end{equation}
Thus the spectrum depends on $D$ and $l$ only via parameter $\mathcal{D}$:
\begin{equation}
E_{nl}(D)=E_{n0}(D+2l)=E_{n0}(\mathcal{D})\,. \label{E-D-l}%
\end{equation}
In the case $D=1$ is the values $l=0$, $1$ correspond the states with positive
and negative parity respectively.

Introducing the variable%
\begin{equation}
\zeta=\frac{r^{2}}{2}%
\end{equation}
we bring $H_{\mathcal{D}}$ (\ref{H-D-def}) to the form%
\begin{equation}
H_{\zeta}=\left(  -\zeta\frac{d^{2}}{d\zeta^{2}}-\frac{\mathcal{D}}{2}%
\frac{d}{d\zeta}\right)  +V(2\zeta)\,. \label{H-zeta}%
\end{equation}
The corresponding Schr\"{o}dinger equation is%
\begin{equation}
\left[  \left(  -\zeta\frac{d^{2}}{d\zeta^{2}}-\frac{\mathcal{D}}{2}%
\frac{d}{d\zeta}\right)  +V(2\zeta)-E_{n}(\mathcal{D})\right]  \psi_{n}%
(\zeta)=0\,. \label{zeta-Schroedinger}%
\end{equation}
Below we assume for simplicity that the potential $V(2\zeta)$ is polynomial in
$\zeta$.

Equation (\ref{zeta-Schroedinger}) was derived for positive integer dimensions
$\mathcal{D}$. However, we can use this equation for analytical continuation
of eigenenergies $E_{n}(\mathcal{D})$ to arbitrary $\mathcal{D}$. To this aim
we solve equation (\ref{zeta-Schroedinger}) imposing the following boundary
conditions on eigenfunctions $\psi_{n}(\zeta)$:

1) $\psi_{n}(\zeta)$ must be analytical at $\zeta=0$:%
\begin{equation}
\phi(\zeta)=\sum\limits_{k=0}^{\infty}p_{k}\zeta^{k}\,, \label{phi-series}%
\end{equation}

2) $\psi_{n}(\zeta)$ must decay at $\zeta\rightarrow+\infty$.

\subsection{$sl(2)$ representation for $O(D)$ symmetric Hamiltonians}

Let us define parameter%
\begin{equation}
j=-\frac{\mathcal{D}}{4} \label{j-j-prime-D}%
\end{equation}
and operators%
\begin{align}
T_{+}  &  =-\zeta\frac{d^{2}}{d\zeta^{2}}+2j\frac{d}{d\zeta}%
,\label{T-plus-diff}\\
T_{0}  &  =-\zeta\frac{d}{d\zeta}+j,\\
T_{-}  &  =\zeta\,. \label{T-minus-diff}%
\end{align}
These operators provide a representation of $sl(2)$ algebra%
\begin{equation}
\lbrack T_{+},T_{-}]=2T_{0},\quad\lbrack T_{+},T_{0}]=-T_{+},\quad\lbrack
T_{-},T_{0}]=T_{-} \label{sl-2-algebra}%
\end{equation}
with the Casimir operator%
\begin{gather}
C_{2}=\frac{1}{2}\left(  T_{+}T_{-}+T_{-}T_{+}\right)  +T_{0}T_{0}\,,\\
\left[  C_{2},T_{a}\right]  =0\,.
\end{gather}
In representation (\ref{T-plus-diff}) -- (\ref{T-minus-diff}) we have%
\begin{equation}
C_{2}=j(j+1)\,. \label{C2-j}%
\end{equation}
In terms of operators (\ref{T-plus-diff}) -- (\ref{T-minus-diff}) Hamiltonian
(\ref{H-zeta}) becomes%
\begin{equation}
H_{T}=T_{+}+V(2T_{-})\,. \label{H-T-def}%
\end{equation}
In the case of the quartic anharmonic oscillator (\ref{H-quartic}) we have%
\begin{equation}
H_{T}^{\mathrm{QO}}=T_{+}+T_{-}+4gT_{-}^{2}\,.
\end{equation}

For integer or half-integer $j$ one could expect a simplification of the
problem of the anharmonic oscillator due to the presence of finite dimensional
irreducible representations of $sl(2)$. However, we meet a problem on this
way: in the case of finite dimensional representations of $sl(2)$ there exists
a vector $\psi$ annihilated by $T_{-}$:%
\begin{equation}
T_{-}\psi=0,\quad\psi\neq0\,,
\end{equation}
whereas for operator $T_{-}$ (\ref{T-minus-diff}) equation%
\begin{equation}
T_{-}\psi(\zeta)=\zeta\psi(z)=0
\end{equation}
leads to $\psi=0$ [or to $\psi(z)=c\delta(z)$ if one allows for generalized
functions]. In the next section we show how this problem can be circumvented
in negative even dimensions.

\subsection{Effective spin Hamiltonian for even negative dimensions
$\mathcal{D}$}

\label{Solution-at-negative-D-section}

Let us consider the case of positive integer or half-integer $j$%
\begin{equation}
j=0,\frac{1}{2},1,\frac{3}{2},\ldots\, \label{j-special}%
\end{equation}
According to eq. (\ref{j-j-prime-D}) this corresponds to negative even values
(\ref{D-even-negative}) of parameter $\mathcal{D}$. These values are
nonphysical. We assume analytical continuation in the sense of solutions of
Schr\"{o}dinger equation with boundary conditions described in Sec.
\ref{Schroedinger-section}.

Let us consider the subspace $\mathcal{H}_{j}$ of functions $f(\zeta)$ obeying
conditions%
\begin{equation}
f^{(k)}(0)=0\quad(0\leq k\leq2j)\,.
\end{equation}
This subspace is invariant under the action of operators $T_{a}$
(\ref{T-plus-diff}) -- (\ref{T-minus-diff}). Therefore we can reduce the
action of operators $T_{a}$ from the space $\mathcal{H}$ of differentiable
functions to the factor space $\mathcal{H}/\mathcal{H}_{j}$. Expression
(\ref{H-T-def}) for Hamiltonian $H_{T}$ shows that $\mathcal{H}_{j}$ is
invariant also under the action of $H_{T}$. Therefore we can also reduce the
action of $H_{T}$ from $\mathcal{H}$ to $\mathcal{H}/\mathcal{H}_{j}$. The
reduction to $\mathcal{H}/\mathcal{H}_{j}$ is also possible for the spectral
problem%
\begin{equation}
\left(  H_{T}-E\right)  \psi=0\,. \label{H-T-Schroedinger}%
\end{equation}
Certainly after this reduction only some finite part of the infinite spectrum
can survive. Note that the factor space $\mathcal{H}/\mathcal{H}_{j}$ is
isomorphic to the space of polynomials $P_{2j}(\zeta)$ of degree $2j$. The
action of operators $T_{a}$ (\ref{T-plus-diff}) -- (\ref{T-minus-diff}) on
these polynomials is described by the usual differentiation algebra extended
by an additional formal rule%
\begin{equation}
\xi^{2j+k}=0\quad\mathrm{for}\quad k\geq1\,.
\end{equation}
The action of operators $T_{a}$ in the space of these polynomials is
equivalent to the spin-$j$ irreducible $(2j+1)$-dimensional representation of
$sl(2)$. This is obvious from the dimension $2j+1$ of this representation and
from the eigenvalue $j(j+1)$ of the Casimir operator (\ref{C2-j}). Therefore
after the reduction of the spectral problem (\ref{H-T-Schroedinger}) to the
factor space $\mathcal{H}/\mathcal{H}_{j}$ we arrive at the matrix eigenvalue
problem%
\begin{gather}
H_{j}=J_{+}+V(2J_{-})\,,\label{H-J-def}\\
\left(  H_{j}-E\right)  \phi=0\, \label{det-E-H-J}%
\end{gather}
where $J_{a}$ are $\left(  2j+1\right)  $-dimensional matrices of the spin-$j$
representation of $sl(2)$. These $sl(2)$ matrices $J_{a}$ are connected by relations%

\begin{align}
J_{0}  &  =J_{3}\,,\\
J_{\pm}  &  =J_{1}\pm iJ_{2} \label{J-pm}%
\end{align}
with standard spin matrices $J_{1,2,3}$ for the spin-$j$ representation of
$su(2)$.

Thus the spectral problem reduces to solving the equation%
\begin{equation}
\det\left(  H_{j}-E\right)  =0\,. \label{H-E-det}%
\end{equation}
Obviously%
\begin{equation}
R_{2j}(E)=\det\left(  H_{j}-E\right)  =\det\left[  J_{+}+V(2J_{-})\,-E\right]
\label{R-2j-def}%
\end{equation}
is a polynomial of degree $2j+1$ so that we deal with an algebraic equation%
\begin{equation}
R_{2j}(E)=0
\end{equation}
where $j=-\mathcal{D}/4$ according to (\ref{j-j-prime-D}).

One should keep in mind that matrix Hamiltonian (\ref{H-J-def}) describes only
a part of the spectrum. Since $sl(2)$ matrices $J_{\pm}$ are real, polynomials
$R_{2j}(E)$ are also real. Nevertheless the eigenvalues of $H_{j} $ may be complex.

In the case of quartic anharmonic oscillator (\ref{H-quartic}) matrix
Hamiltonian (\ref{H-J-def}) reduces to%
\begin{equation}
H_{j}=J_{+}+J_{-}+4gJ_{-}^{2}\,. \label{H-J-quartic}%
\end{equation}

\subsection{Connection with quasi-exactly solvable problems}

The Hamiltonians of type (\ref{H-J-quartic}) appear in the context of
quasi-exactly solvable (QES) potentials
\cite{Turbiner-94,Shifman-ITEP,Ushveridze}. However, there is a certain
difference between the $sl(2)$ representation discussed here and in
traditional QES problems. Our differential operators $T_{a}$
(\ref{T-plus-diff}) -- (\ref{T-minus-diff}) differ from the operators
$T_{a}^{\prime}$ used in QES problems:%
\begin{align}
T_{+}^{\prime}  &  =2j\xi-\xi^{2}\frac{d}{d\xi}\,,\\
T_{0}^{\prime}  &  =-j+\xi\frac{d}{d\xi}\,,\\
T_{-}^{\prime}  &  =\frac{d}{d\xi}\,.
\end{align}
Both sets of differential operators obey $sl(2)$ algebra (\ref{sl-2-algebra}).
In fact, $T_{a}$ and $T_{a}^{\prime}$ are connected (at least formally) by a
Laplace transformation combined with the change of $j\rightarrow-1-j$.
Nevertheless the two classes of problems are different. In particular, the
methods discussed here allow us to compute a part of the spectrum but not the
corresponding wave functions $\psi_{n}(\zeta)$ obeying equation
(\ref{zeta-Schroedinger}). It should be emphasized that one has to distinguish
two types of wave functions at negative $\mathcal{D}$:

1) solutions of Schr\"{o}dinger equation (\ref{zeta-Schroedinger}) at negative
even $\mathcal{D}$,

2) eigenvectors of the effective spin Hamiltonian (\ref{H-J-def}).

It is interesting that in the case of the sextic QES potential, Dunne and
Halliday \cite{DH-88} could trace the connection between the exact
eigenfunctions in physical dimensions $D=1,2,3,\ldots$ and the eigenvectors of
the effective finite-dimensional matrix problem corresponding to
$D=0,-2,-4\ldots$

\subsection{Example: harmonic oscillator}

In the case of the harmonic oscillator%
\begin{align}
V^{\mathrm{HO}}(r^{2})  &  =\frac{1}{2}r^{2}\,,\label{V-HO}\\
V^{\mathrm{HO}}(2\zeta)  &  =\zeta\,,
\end{align}
the calculation of determinant (\ref{det-E-H-J}) is trivial:%
\begin{equation}
\det\left(  J_{+}+J_{-}-E\right)  =\det\left(  2J_{1}-E\right)  =\det\left(
2J_{3}-E\right)  =2^{2j+1}\prod\limits_{n=-j}^{j}\left(  n-\frac{E}{2}\right)
\label{det-H-H-HO}%
\end{equation}
so that solutions of eq. (\ref{H-E-det}) are%
\begin{align}
E  &  =-2j,-2\left(  j-1\right)  ,\ldots,2\left(  j-1\right)  ,2j\nonumber\\
&  =\frac{\mathcal{D}}{2},\frac{\mathcal{D}}{2}+2,\ldots,-\frac{\mathcal{D}%
}{2}-2,-\frac{\mathcal{D}}{2}\quad(\mathcal{D}\leq0). \label{E-matrix-EV-HO}%
\end{align}
It is instructive to compare this algebraic solution with the full spectrum of
the harmonic oscillator. Labeling levels with $n=0,1,2,3,\ldots$ in each $l$
sector, we can write%
\begin{equation}
E_{nl}(0,D)=\frac{D}{2}+\left(  2n+l\right)  \,. \label{E-nl-HO}%
\end{equation}
Combining $D$ and $l$ into parameter $\mathcal{D}$ (\ref{Delta-def}) we arrive
at the expression%
\begin{equation}
E=\frac{\mathcal{D}}{2}+2n,\quad n=0,1,2,\ldots\label{HO-all-levels}%
\end{equation}
that has a trivial analytical continuation in $\mathcal{D}$. For negative even
$\mathcal{D}=-4j$ the lowest $2j+1$ levels (\ref{HO-all-levels}) obviously
coincide with the solution of the spin problem (\ref{E-matrix-EV-HO}).

\begin{figure}[ptb]
\begin{center}
\includegraphics[
height=2.1015in,
width=3.2958in
]{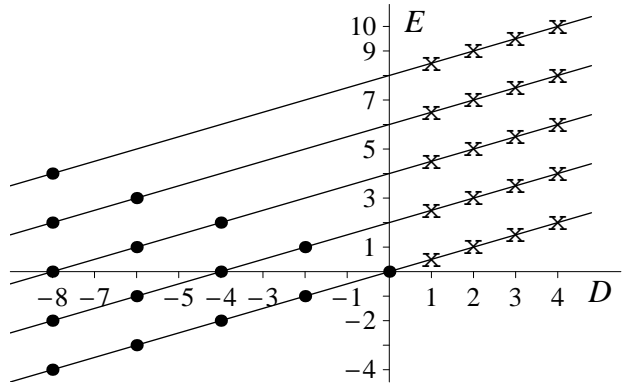}
\end{center}
\caption{Trajectories continuing the levels of the harmonic oscillator
analytically in $\mathcal{D}$ from physical values $\mathcal{D}=1,2,3,\ldots$
(marked with crosses) to arbitrary values of $\mathcal{D}$. The levels at
$\mathcal{D}=0,-2,-4,\ldots$ that are described by roots of polynomials
$R_{2j}$ ($j=-D/2$) are marked with circles.}%
\label{trajectories-HO-fig}%
\end{figure}

In Fig. \ref{trajectories-HO-fig} we show the ``Regge trajectories'' (the
relation between the analytical continuation in $D$ and Regge theory is
discussed in Sec. \ref{Regge-theory-subsection}) connecting physical values
$D=1,2,3,\ldots$ and with the special points $D=0,-2,-4,\ldots$ It is
remarkable that any trajectory corresponding to a physical level after
analytical continuation to negative $D$ sooner or later enters into the domain
described by the roots of polynomial (\ref{det-H-H-HO}).

\subsection{Power series and recursion relations}

\label{Recusrion-relation-subsection}

The analytical continuation of levels $E_{n}(\mathcal{D})$ to negative
$\mathcal{D}$ uses solutions $\phi(\zeta)$ (\ref{phi-series}) of
Schr\"{o}dinger equation (\ref{zeta-Schroedinger}) that are analytical at
$\zeta=0$. For polynomial potentials%
\begin{align}
V(2\zeta)  &  =\sum\limits_{k}w_{k}\zeta^{k}\,,\label{V-w}\\
w_{k}  &  =0\quad\mathrm{if}\quad k\leq0\quad\mathrm{or}\quad k>L
\end{align}
obeying condition%
\begin{equation}
V(0)=0
\end{equation}
let us define%
\begin{equation}
u_{k}=\left\{
\begin{array}
[c]{ll}%
w_{k} & \mathrm{if}\quad\,1\leq k\leq L,\\
-E & \mathrm{if}\quad k=0,\\
0 & \mathrm{otherwise\,.}%
\end{array}
\right.  \label{u-k-def}%
\end{equation}
Then%
\begin{equation}
V(2\zeta)-E=\sum\limits_{k=-\infty}^{\infty}u_{k}\zeta^{k}\,.
\end{equation}
Inserting this series and expansion (\ref{phi-series}) into eq.
(\ref{zeta-Schroedinger}) one obtains the recursion relation for the
coefficients $p_{k}$ of series (\ref{phi-series}):%
\begin{align}
a_{k+1}(\mathcal{D})p_{k+1}  &  =\sum\limits_{m=0}^{L}u_{m}p_{k-m}%
\,\,,\label{recursion-0}\\
a_{k}(\mathcal{D})  &  =k\left(  k-1+\frac{1}{2}\mathcal{D}\right)  \,.
\label{a-k-D-def}%
\end{align}
If $D\neq0,-2,-4,\ldots$ then we can solve this equation iteratively starting
from some $p_{0}\neq0$ and obtain%
\begin{equation}
p_{k+1}=\frac{p_{0}s_{k+1}\left(  \{u_{m}\},\mathcal{D}\right)  }%
{a_{k+1}(\mathcal{D})} \label{p-s}%
\end{equation}
where $s_{k+1}\left(  \{u_{m}\},\mathcal{D}\right)  $ are polynomials of
$u_{m}$. Once the series (\ref{phi-series}) is constructed, one has impose the
boundary condition on $\phi(\zeta)$ at $\zeta\rightarrow+\infty$ and this will
fix the spectrum.

The case $\mathcal{D}\equiv-4j=0,-2,-4,\ldots$ is exceptional because in this
case at $k=-\mathcal{D}/2=2j$ we have a zero in the denominator on the RHS of
(\ref{p-s}). Therefore instead of the determination of $p_{2j}$ we will arrive
at the condition%
\begin{equation}
s_{2j+1}\left(  \{u_{m}\},-4j\right)  =0\,. \label{s-condition}%
\end{equation}
Since $s_{2j+1}\left(  \{u_{m}\},\mathcal{D}\right)  $ is a nonzero polynomial
in $E$ (via $u_{0}=-E$), this condition does not hold for the most of
energies. In fact, in the case $\mathcal{D}=0,-2,-4,\ldots$ the majority of
eigenfunctions have power expansion (\ref{phi-series}) with%
\begin{equation}
p_{0}=p_{1}\ldots=p_{2j+1}=0,\quad p_{2j+2}\neq0\,. \label{p-type-2}%
\end{equation}

However, if condition (\ref{s-condition}) holds then we can proceed with
iterations and construct the series (\ref{phi-series}). Thus in the special
case (\ref{s-condition}) we have two solutions (\ref{phi-series}) of
Schr\"{o}dinger equation (\ref{zeta-Schroedinger}) analytical at $\zeta=0$:
one solution with $p_{0}\neq1$ and the second solution (\ref{p-type-2}). Thus
all solutions of Schr\"{o}dinger equation (\ref{phi-series}) are regular in
the case (\ref{s-condition}). This means that taking an appropriate linear
combination we can satisfy boundary conditions at infinity so that values of
$E$ obeying equation (\ref{s-condition}) automatically belong to the spectrum.

It is easy to see that this construction is equivalent to the spin problem
(\ref{H-E-det}). Indeed, recursion relation (\ref{recursion-0}) is nothing
else but vector equation (\ref{H-E-det}) written in the basis diagonalizing
matrix $J_{0}$. Now we understand that polynomials $s_{2j+1}\left(
\{w_{m}\},E,-4j\right)  $ appearing in eq. (\ref{s-condition}) must coincide
up to a constant with polynomials $R_{2j}(E)$ defined by eq. (\ref{R-2j-def})%
\begin{equation}
s_{2j+1}\left(  \{u_{m}\},-4j\right)  =c_{j}\det\left[  J_{+}+V\left(
2J_{-}\right)  -E\right]  \label{s-det-equality}%
\end{equation}
with some coefficients $c_{j}$.

Let us derive equation (\ref{s-det-equality}) more carefully and compute
coefficients $c_{j}$. First we define a $\left(  N+1\right)  \times\left(
N+1\right)  $ matrix%
\begin{equation}
C_{kn}^{(N)}=u_{k-n}-a_{k}\delta_{k,n-1}\quad1\leq k,n\leq N+1\,.
\label{C-N-def}%
\end{equation}
According to eq. (\ref{u-k-def}) we assume that $u_{k}=0$ if $k<0$.

Then recursion relation (\ref{recursion-0}) takes the form%
\begin{equation}
\sum\limits_{n=1}^{N+1}C_{kn}^{(N)}p_{n-1}=\delta_{k,N+1}a_{N+1}p_{N+1}%
\quad(1\leq k\leq N+1)\,. \label{C-recursion}%
\end{equation}
Matrix (\ref{C-N-def}) is quasitriangular. Its determinant can be computed by
making with its rows the same linear manipulations as in the recursive
calculation of $s_{k+1}\left(  \{u_{m}\},\mathcal{D}\right)  $ in eq.
(\ref{p-s}). As a result, one obtains%
\begin{equation}
\det\left[  C^{(N)}\left(  \{u_{m}\},\mathcal{D}\right)  \right]
=s_{N+1}\left(  \{u_{m}\},\mathcal{D}\right)  \prod\limits_{k=1}^{N}%
a_{k}(\mathcal{D})\,.
\end{equation}
Now we set $\mathcal{D}=-4j,$ $N=2j$. Using eq. (\ref{a-k-D-def}), we find%
\begin{equation}
\prod\limits_{k=1}^{2j}a_{k}(-2j)=(-1)^{2j}\left[  (2j)!\right]  ^{2}\,.
\end{equation}
Thus%
\begin{equation}
\det\left[  C^{(2j)}\left(  \{u_{m}\},-4j\right)  \right]  =(-1)^{2j}\left[
(2j)!\right]  s_{2j+1}\left(  \{u_{m}\},-4j\right)  \,. \label{det-C-s}%
\end{equation}
Therefore algebraic equation (\ref{s-condition}) takes the form%
\begin{equation}
\det\left[  C^{(2j)}\left(  \{u_{m}\},-4j\right)  \right]  =0\,.
\label{det-C-0}%
\end{equation}
In practical calculations one can compute det$C^{(2j)}$ either directly using
definition (\ref{C-N-def})%
\begin{equation}
\det C^{(2j)}=\det_{1\leq k,n\leq2j+1}\left[  w_{k-n}-k\left(  k-1-2j\right)
\delta_{k,n-1}-E\delta_{kn}\right]  \label{det-C-via-w}%
\end{equation}
or solving recursion relations iteratively and finding in this way
$s_{2j+1}\left(  \{u_{m}\},-4j\right)  $. Both methods lead to the same
results:%
\begin{align}
\det C^{(0)}  &  =-E\,,\\
\det C^{(1)}  &  =E^{2}-w_{1}\,,\\
\det C^{(2)}  &  =-E^{3}+4w_{2}+4Ew_{1}\,,\\
\det C^{(3)}  &  =E^{4}-10E^{2}w_{1}+9w_{1}^{2}-24w_{2}E-36w_{3}\,,\\
\det C^{(4)}  &  =-E^{5}+20E^{3}w_{1}-64Ew_{1}^{2}+84w_{2}E^{2}-192w_{1}%
w_{2}+288Ew_{3}+576w_{4}\,.
\end{align}

Note that recursion relation (\ref{C-recursion}) for $N=2j$ is nothing else
but the equation%
\begin{equation}
\left[  T_{+}+V\left(  2T_{-}\right)  -E\right]  \phi(\zeta)=0
\end{equation}
written in the basis $\zeta^{k}$, $k=0,1,2,\ldots2j$. This basis coincides
with the standard basis for spin-$j$ representation up to the subtleties of
the normalization and enumeration. Obviously the determinant of matrix
$C^{(2j)}$ is invariant with the respect to this trivial change of the basis.
Therefore%
\begin{equation}
\det C^{(2j)}=\det\left[  J_{+}+V\left(  2J_{-}\right)  -E\right]
\label{det-C-det-H-J}%
\end{equation}
where the spin $j$ representation is assumed on the RHS. Combining this result
with eq. (\ref{det-C-0}) we see that we have obtained old equation
(\ref{H-J-def}).

Certainly this ``new derivation'' of eq. (\ref{H-J-def}) is nothing else but
an explicit detailed version of the compact arguments of Sec.
\ref{Solution-at-negative-D-section} presented there in terms of the $sl(2)$
algebra acting in the factor space $\mathcal{H}/\mathcal{H}_{j}$.

\subsection{Properties of polynomials $R_{2j}(E)$}

Comparing eqs. (\ref{R-2j-def}), (\ref{det-C-via-w}), (\ref{det-C-det-H-J}) we
see that we have three equivalent representations for polynomials $R_{2j}(E)$%
\begin{align}
R_{2j}(E)  &  =\det\left[  J_{+}+V(2J_{-})\,-E\right] \label{R-repr-1}\\
&  =\det_{1\leq k,n\leq2j+1}\left[  w_{k-n}-k\left(  k-1-2j\right)
\delta_{k,n-1}-E\delta_{kn}\right] \label{R-repr-2}\\
&  =(-1)^{2j}\left[  (2j)!\right]  s_{2j+1}\left(  \{u_{m}\},-4j\right)  \,.
\label{R-repr-3}%
\end{align}
Representation (\ref{R-repr-1}) is written in terms of spin-$j$ matrices
$J_{\pm}=J_{1}\pm iJ_{2}$ and explicitly expresses the $sl(2)$ symmetry
standing behind this construction. Representation (\ref{R-repr-2}) written in
terms of coefficients $w_{k}$ of the polynomial potential $V$ (\ref{V-w}) is
nothing else but eq. (\ref{R-repr-1}) transformed to a basis with a
nonstandard normalization which allows us to get rid of cumbersome square
roots appearing in the traditional expressions for the matrix elements of
$J_{\pm}$. Representation (\ref{R-repr-3}) is based on the iterative solution
(\ref{p-s}) of recursion relations (\ref{recursion-0}) and is useful for
computer calculations of polynomials $R_{2j}(E)$.

The higher coefficients of polynomials $R_{2j}(E)$ can be computed using the
large-$E$ expansion:%
\begin{gather}
\det\left[  E-J_{+}-V(2J_{-})\right]  =E^{2j+1}\exp\left\{  \mathrm{Tr}%
\ln\left[  1-\frac{J_{+}+V(2J_{-})}{E}\right]  \right\} \nonumber\\
=E^{2j+1}\exp\left\{  -\sum\limits_{n=1}^{\infty}\frac{1}{n}E^{-n}%
\mathrm{Tr}\left[  J_{+}+V(2J_{-})\right]  ^{n}\right\}  \,.
\end{gather}
We have%
\begin{align}
\mathrm{Tr}\left[  J_{+}+V(2J_{-})\right]   &  =0\,,\nonumber\\
\mathrm{Tr}\left\{  \left[  J_{+}+V(2J_{-})\right]  ^{2}\right\}   &
=2\mathrm{Tr}\left[  J_{+}V(2J_{-})\right]  =2w_{1}\mathrm{Tr}\left(
J_{+}J_{-}\right)  \,,\nonumber\\
\mathrm{Tr}\left[  J_{+}+V(2J_{-})\right]  ^{3}  &  =3\mathrm{Tr}\left[
\left(  J_{+}\right)  ^{2}V(2J_{-})\right]  =3w_{2}\mathrm{Tr}\left[  \left(
J_{+}\right)  ^{2}\left(  J_{-}\right)  ^{2}\right]
\end{align}
where $w_{k}$ are coefficients of the polynomial potential (\ref{V-w}). A
straightforward spin algebra gives%
\begin{align}
\mathrm{Tr}\left(  J_{+}J_{-}\right)   &  =\frac{2}{3}j(j+1)(2j+1)\,,\\
\mathrm{Tr}\left[  \left(  J_{+}\right)  ^{2}\left(  J_{-}\right)
^{2}\right]   &  =\frac{2}{15}j(j+1)(2j+1)(2j-1)(2j+3)\,.
\end{align}
Finally we obtain%

\begin{align}
(-1)^{2j+1}R_{2j}(E)  &  =E^{2j+1}-\frac{2}{3}j(j+1)(2j+1)w_{1}E^{2j-1}%
\nonumber\\
&  -\frac{2}{15}j(j+1)(2j+1)(2j-1)(2j+3)w_{2}E^{2j-2}+O(E^{2j-3}),
\end{align}

In the case of harmonic oscillator (\ref{V-HO}) we have according to
(\ref{det-H-H-HO})%
\begin{equation}
R_{2j}^{\mathrm{HO}}(E)=\det\left(  J_{+}+J_{-}-E\right)  =2^{2j+1}%
\prod\limits_{n=-j}^{j}\left(  n-\frac{E}{2}\right)  \,.
\end{equation}

\subsection{Symmetry $D\rightarrow4-D$}

\label{D-4-D-symmetry-section}

Relation%

\begin{equation}
\zeta^{-1+(\mathcal{D}/2)}\left(  -\zeta\frac{d^{2}}{d\zeta^{2}}%
-\frac{\mathcal{D}}{2}\frac{d}{d\zeta}\right)  \zeta^{1-(\mathcal{D}%
/2)}=-\zeta\frac{d^{2}}{d\zeta^{2}}-\frac{4-\mathcal{D}}{2}\frac{d}{d\zeta}
\label{D-4-D-symmetry}%
\end{equation}
shows that equation (\ref{zeta-Schroedinger}) has a symmetry%
\begin{equation}
\phi_{4-\mathcal{D}}(\zeta)=\zeta^{-1+(\mathcal{D}/2)}\phi_{\mathcal{D}}%
(\zeta)\,, \label{D-4-D-symmetry-phi}%
\end{equation}%
\begin{equation}
\mathcal{D}\rightarrow4-\mathcal{D\,}\,.
\end{equation}

However, one should be careful about the boundary conditions. Indeed, for
$\mathcal{D}<2$ the factor $\zeta^{-1+(\mathcal{D}/2)}$ is singular so that
regular solutions $\phi_{\mathcal{D}}(\zeta)$ may correspond to singular
solutions of $\phi_{4-\mathcal{D}}(\zeta)$. In other words, some part of the
$\mathcal{D}<2$ spectrum may be lost when one turns from $\mathcal{D}<2$ to
$4-\mathcal{D}>2$.

According to results of Sec. \ref{Recusrion-relation-subsection} at
$\mathcal{D}=2M=0,-2,-4,\ldots$ there are two types of regular solutions
(\ref{phi-series})

1) solutions with $p_{0}\neq0$,

2) solutions with $p_{0}=\ldots=p_{M}=0$, $p_{M+1}\neq0$.

Obviously solutions $\phi_{\mathcal{D}}(\zeta)$ of the first type with
$p_{0}\neq0$ generate singular functions $\phi_{4-\mathcal{D}}(\zeta)$
(\ref{D-4-D-symmetry-phi}). Therefore the spectrum of the $4-\mathcal{D}$
problem does not contain that part of the $\mathcal{D}=-0,-2,-4$ spectrum
which corresponds to wave functions (\ref{phi-series}) with $p_{0}=0$. We see
that under the change $\mathcal{D}\rightarrow4-\mathcal{D}$ we lose exactly
those states which are described by the roots of polynomials $R_{M}(E)$.

This symmetry $\mathcal{D}\rightarrow4-\mathcal{D}$ is illustrated in Fig.
\ref{trajectories-AO-fig}. It should be stressed that the symmetry
$\mathcal{D}\rightarrow4-\mathcal{D}$ is relevant only for integer even values
of $\mathcal{D}$.

\section{Quartic anharmonic oscillator}

\setcounter{equation}{0} 

\subsection{Polynomials $R_{2j}(E,g)$}

In the case of the quartic anharmonic oscillator (\ref{H-quartic}) we have%
\begin{align}
V(2\zeta)  &  =\zeta+4g\zeta^{2}=w_{1}+w_{2}\zeta^{2}\,,\\
w_{1}  &  =1\,,\quad w_{2}=4g\,.
\end{align}
Now one can use one of representations (\ref{R-repr-1}) -- (\ref{R-repr-3})
and compute $R_{2j}(E)$. In order to avoid large coefficients it is convenient
to define%

\begin{equation}
\tilde{R}_{2j}(E,g)=(-2)^{-2j}\mathrm{\,}R_{2j}(E)=(-2)^{-2j}\mathrm{\,}%
\det\left(  J_{+}+J_{-}+4gJ_{-}^{2}-E\right)  \,. \label{R-tilde-2j-def}%
\end{equation}
At $g=0$ we have according to (\ref{det-H-H-HO})%
\begin{equation}
\tilde{R}_{2j}(E,0)=-2\prod_{k=0}^{2j}\left(  \frac{E-2j}{2}+k\right)  \,.
\label{R-HO-res}%
\end{equation}
Let us define%
\begin{equation}
\tilde{R}_{2j}^{\prime}(E,g)=\tilde{R}_{2j}(E,g)-\tilde{R}_{2j}(E,0)\,.
\label{R-prime-def}%
\end{equation}
The first polynomials $\tilde{R}_{2j}^{\prime}(E,g)$ are%
\begin{align}
\tilde{R}_{1}^{\prime}(E,g)  &  =0\,,\\
\tilde{R}_{2}^{\prime}(E,g)  &  =4g\,,\\
\tilde{R}_{3}^{\prime}(E,g)  &  =12Eg\,,\\
\tilde{R}_{4}^{\prime}(E,g)  &  =3(-16+7E^{2})g\,,\\
\tilde{R}_{5}^{\prime}(E,g)  &  =4(7E^{3}-55E-200g)g\,,\\
\tilde{R}_{6}^{\prime}(E,g)  &  =\frac{9}{2}(7E^{4}-124E^{2}+192-1000Eg)g\,,\\
\tilde{R}_{7}^{\prime}(E,g)  &  =\frac{9}{2}(7E^{5}-230E^{3}+1183E-3056E^{2}%
g+10976g)g\,.
\end{align}
Now we find using eqs. (\ref{R-HO-res}) and (\ref{R-prime-def})%
\begin{align}
\tilde{R}_{0}(E,g)  &  =-E\,,\label{R-0}\\
\tilde{R}_{1}(E,g)  &  =-\frac{1}{2}(E^{2}-1)\,,\label{R-1}\\
\tilde{R}_{2}(E,g)  &  =-\frac{1}{4}E(E^{2}-4)+4g\,,\label{R-2}\\
\tilde{R}_{3}(E,g)  &  =-\frac{1}{8}(E^{2}-9)(E^{2}-1)+12gE. \label{R-3}%
\end{align}

\subsection{Ground state at $D=-4$}

The ground state has $l=0$ so that the case of the ground state in $D=-4$
dimensions corresponds to $\mathcal{D}=D=-4$. The spectrum is described by the
roots of polynomial $\tilde{R}_{2}(E,g)$ (\ref{R-2})%
\begin{equation}
\tilde{R}_{2}(E,g)=0
\end{equation}
i.e.%
\[
E^{3}-4E-16g=0\,.
\]
Let us define%
\begin{align}
E  &  =\frac{2}{\sqrt{3}}\varepsilon\,,\\
h  &  =3^{3/2}g\,.
\end{align}
Then%
\begin{equation}
\varepsilon^{3}-3\varepsilon-2h=0\,.
\end{equation}
The solutions can be found using Cardano formula:%
\begin{equation}
\varepsilon=\sqrt[3]{h+\sqrt{h^{2}-1}}+\sqrt[3]{h-\sqrt{h^{2}-1}}\,.
\end{equation}
Keeping in mind applications to the case of the perturbation theory in small
$g$ it is convenient to transform this expression to the form%
\begin{align}
\varepsilon &  =e^{5\pi i/6}\sqrt[3]{\sqrt{1-h^{2}}-ih}+e^{-5\pi i/6}%
\sqrt[3]{\sqrt{1-h^{2}}-ih}\nonumber\\
&  =2\mathrm{Re}\left[  \left(  \frac{-\sqrt{3}+i}{2}\right)  \sqrt[3]%
{\sqrt{1-h^{2}}-ih}\right]
\end{align}
which allows for the perturbative expansion%
\begin{equation}
\varepsilon=-\sqrt{3}+\frac{h}{3}+\ldots
\end{equation}
This leads to the perturbative expansion for the energy of the ground state in
$D=-4$ dimensions%
\begin{equation}
E(g,-4)=-2+2g+3g^{2}+8g^{3}+\frac{105}{4}g^{4}+96g^{5}+\frac{3003}{16}%
g^{6}+1536g^{7}+\ldots\label{E-D-4-series}%
\end{equation}
In Sec. \ref{Higher-orders-section} we will compute the asymptotic behavior
(\ref{E-k-4-asymptotic}) of this series.

\section{Analytical continuation to negative $D$}

\setcounter{equation}{0} 

\subsection{Negative $\mathcal{D}$ and Regge theory}

\label{Regge-theory-subsection}

We have already made some comments about the role of Schr\"{o}dinger equation
(\ref{zeta-Schroedinger}) for the analytical continuation in $\mathcal{D}$.
Remember that parameter $\mathcal{D}$ (\ref{Delta-def}) is built of $D$ and
$l$ so that the problem of analytical continuation in $\mathcal{D}$ has two
equivalent formulations:

1) analytical continuation in $D$ at fixed $l$,

2) analytical continuation in $l$ at fixed $D$.

The second approach allows us to use some results from Regge theory of complex
angular momenta \cite{DeAlfaro-Regge,Collins-book,Newton-book}. Although the
traditional Regge theory deals with the scattering problem whereas we are
interested in polynomial potentials, the analysis of the behavior of the
solutions of Schr\"{o}dinger equation (\ref{zeta-Schroedinger}) at
$\zeta\rightarrow0$ is the same in Regge theory of the potential scattering
and in our case. In particular, in Regge theory for potentials regular at
$r=0$, the Regge poles are at points%
\begin{equation}
l=-\frac{3}{2},-\frac{5}{2},-\frac{7}{2}\ldots\label{l-Regge}%
\end{equation}
Combining this with the general formula (\ref{Delta-def}) applied to the case $D=3$%

\begin{equation}
\mathcal{D}=3+2l\quad(D=3) \label{D-l-dependence-D3}%
\end{equation}
we see that $l$-points (\ref{l-Regge}) correspond to negative even values
$\mathcal{D}=0,-2,-4,-6,\ldots$.

From the point of view of the historical perspective it is also interesting
that analytical continuation in $l$ for potentials growing at infinity, e.g.
for the oscillator energies (\ref{E-nl-HO}), was discussed in the context of
naive models for quark confinement \cite{Collins-book,Newton-book}.

Although the discussion of analytical continuation in terms of $l$ is
preferable from the point of view of explicit connections with Regge theory,
we choose the $D$ representation. The $D$ language is natural for the analysis
of analytical properties of the partition function (Sec.
\ref{Analytical-PF-section}) and for the fermion representation at negative
$D$ (\ref{Fermion-section}).

\subsection{$D$ dependence of levels}

\begin{figure}[ptb]
\begin{center}
\includegraphics[
height=2.1041in,
width=3.3235in
]{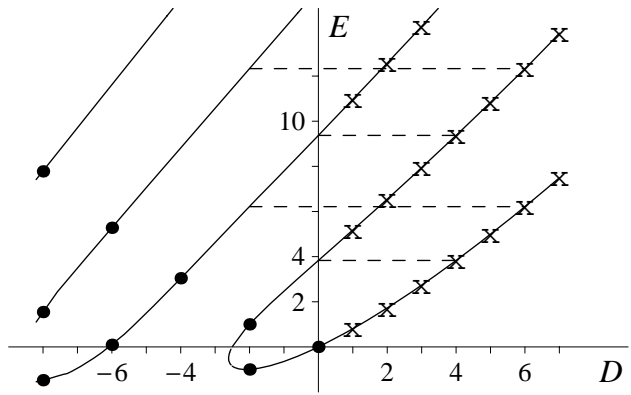}
\end{center}
\caption{Trajectories continuing the levels of the quartic anharmonic
oscillator with $g=1$ analytically in $\mathcal{D}$ from physical values
$\mathcal{D}=1,2,3,\ldots$ (marked with crosses) to arbitrary values of
$\mathcal{D}$. The levels at $\mathcal{D}=0,-2,-4,\ldots$ that are described
by roots of polynomials $R_{2j}$ ($j=-D/2$) are marked with circles. Note that
the trajectories of the ground state and of the first excited state meet
together at $D\approx-2.6$ . Some polynomials $R_{2j}$ have complex roots so
that the number of real levels associated with these polynomials is smaller
than $2j+1$. The dashed horizontal lines connect levels related by the
symmetry $D\rightarrow4-D$ (for even $D$).}%
\label{trajectories-AO-fig}%
\end{figure}

We have already discussed the dependence of energy $E_{n}(\mathcal{D})$ on
$\mathcal{D}$ for the harmonic oscillator (see Fig. \ref{trajectories-HO-fig}%
). If we switch on a small anharmonicity then the trajectories will be
deformed but qualitative features will be preserved. However, in the case of a
strong deviation from the harmonic regime (or at larger values of
$|\mathcal{D}|$) some new phenomena may occur. In Fig.
\ref{trajectories-AO-fig} we show the $\mathcal{D}$ dependence of levels for
the quartic anharmonic oscillator (\ref{H-quartic}) with $g=1$. We see that
the trajectories of the ground state and of the first excited state meet
together at $\mathcal{D}\approx-2.6$ . Polynomials $R_{2j}$ have $2j+1$ roots
but some of these roots may be complex. In Fig. \ref{trajectories-AO-fig} we
see that this happens for polynomials $R_{2}$, $R_{3}$, $R_{4}$.

%EndExpansion

\subsection{Various aspects of the problem of analytical continuation in
$\mathcal{D}$}

\label{Various-analytical-aspects}

The problem of analytical continuation in $\mathcal{D}$ can be studied for
various quantities:

1) analytical continuation of separate energy levels
\begin{equation}
E_{nl}(D)=E_{n0}(D+2l)=E_{n0}(\mathcal{D})\,,
\end{equation}
.

2) analytical continuation of coefficients $E_{n}^{(k)}(\mathcal{D})$ of the
perturbative expansion%
\begin{equation}
E_{n0}(\mathcal{D},g)=\sum\limits_{k=1}^{\infty}E_{n0}^{(k)}(\mathcal{D})g^{k}
\label{E-n-g-epansion}%
\end{equation}
for anharmonic oscillator,

3) analytical continuation of the partition function%
\begin{equation}
Z(\beta,D)=\mathrm{Tr}\,\exp\left(  -\beta H\right)  \,. \label{Z-beta-def}%
\end{equation}
Let us comment briefly on the analytical properties of these quantities.

The situation with analytical continuation of the perturbative coefficients
$E_{n0}^{(k)}(\mathcal{D})$ in expansion (\ref{E-n-g-epansion}) is simple
because $E_{n0}^{(k)}(\mathcal{D})$ are polynomials of $\mathcal{D}$. However,
the asymptotic nature of the perturbative series (\ref{E-n-g-epansion}) does
not allow us to draw simple conclusions about the analyticity in $\mathcal{D}$
of exact levels $E_{n0}(\mathcal{D},g)$ from the trivial analyticity of
$E_{n0}^{(k)}(\mathcal{D}).$

The simplest example of analytical continuation in $D$ provides the harmonic
oscillator. Its partition function trivially factorizes into $D$ components%
\begin{equation}
Z^{(0)}(\beta,D,0)=\left[  Z^{(0)}(\beta,1,0)\right]  ^{D}%
\end{equation}
with an obvious analyticity in $D$. The case of anharmonic potentials will be
considered in the next section.

\subsection{Analytical continuation of the partition function}

\label{Analytical-PF-section}

Now let us study analytical continuation of the partition function%
\begin{equation}
Z(\beta,D)=\mathrm{Tr}\exp\left(  -\beta H\right)  \quad(D=1,2,3,\ldots)
\end{equation}
in $D$. At physical values of $D=1,2,3,4,\ldots$ one can use the $l$
decomposition%
\begin{equation}
Z(\beta,D)=\sum\limits_{n,l=0}^{\infty}m(D,l)\exp\left[  -\beta E_{nl}%
(D)\right]  \,. \label{Z-D-l-sum-1}%
\end{equation}
Here $m(D,l)$ is the degeneracy of the $l$-levels in the $D$-dimensional space
given by eq. (\ref{l-degeneracy-positive}). Using eq. (\ref{E-D-l}), we find
from eq. (\ref{Z-D-l-sum-1})%

\begin{equation}
Z(\beta,D)=\sum\limits_{l=0}^{\infty}m(D,l)\left\{  \sum_{n}\exp\left[  -\beta
E_{n0}(D+2l)\right]  \right\}  \,.
\end{equation}
Let us define%
\begin{equation}
z(\beta,D)=\sum\limits_{n=0}^{\infty}\exp\left[  -\beta E_{n0}(D)\right]  \,.
\end{equation}
Then%
\begin{equation}
Z(\beta,D)=\sum\limits_{l=0}^{\infty}m(D,l)\,z(\beta,D+2l)\,.
\label{Z-D-l-sum-2}%
\end{equation}
We want to use this series for analytical continuation of $Z(\beta,D)$ to
negative $D$. It is easy to check that for positive integer $l=0,1,2,3\ldots$,
function $m(D,l)$ is analytical in $D$ in the whole complex plane. Analytical
continuation of $m(D,l)$ to $D=1$ and $D=2$ was already discussed in Sec.
\ref{Schroedinger-section}. Now let us consider the general case. Using the
relation%
\begin{equation}
\Gamma(z)\Gamma(1-z)=\frac{\pi}{\sin\pi z}\,,
\end{equation}
we can transform eq. (\ref{l-degeneracy-positive}) to the form%

\begin{equation}
m(D,l)=(-1)^{l}\frac{(2l+D-2)}{l!}\frac{\Gamma(2-D)}{\Gamma(3-D-l)}\,.
\label{m-D-l-2}%
\end{equation}
Combining representations (\ref{l-degeneracy-positive}) and (\ref{m-D-l-2}),
we see that $m(D,l)$ is regular for all $D$. The regularity of $m(D,l)$ allows
us to use representation (\ref{Z-D-l-sum-2}) for analytical continuation of
$Z(D)$ to negative $D$.

At negative even values $D=-2M=0,-2,-4,\ldots$ we have an important
simplification. Indeed, according to eq. (\ref{m-D-l-2})%
\begin{equation}
m(-2M,l)=(-1)^{l+1}\frac{(2l-2M-2)}{l!}\frac{\Gamma(2+2M)}{\Gamma(3+2M-l)}\,.
\end{equation}
We see that%
\begin{equation}
m(-2M,l)=0\quad\mathrm{if}\quad l\geq2M+3\,.
\end{equation}
Therefore series (\ref{Z-D-l-sum-2}) reduces to a finite sum:%
\begin{equation}
Z(\beta,-2M)=\sum\limits_{l=0}^{2(M+1)}(-1)^{l}\frac{2(-l+M+1)}{l!}%
\frac{(1+2M)!}{(2+2M-l)!}\,\,z(\beta,-2M+2l)\,.
\end{equation}
Now we change the summation variable from $l$ to%
\begin{equation}
k=l-M-1\,.
\end{equation}
Then%
\begin{equation}
Z(\beta,-2M)=\sum\limits_{k=-M-1}^{M+1}\frac{(-1)^{k-M}2k\left[
(1+2M)!\right]  }{(M+1-k)!(1+M+k)!}\,\,z(\beta,2+2k)\,.
\end{equation}
Note that $k=0$ does not contribute to this sum. Combining the contributions
of $k$ and $-k$ we arrive at%
\begin{align}
Z(\beta,-2M)  &  =\sum\limits_{k=1}^{M+1}\frac{(-1)^{k-M}2k\left[
(1+2M)!\right]  }{(M+1-k)!(1+M+k)!}\nonumber\\
&  \times\,\,\left[  z(\beta,2+2k)-z(\beta,2-2k)\right]  \,.
\label{Z-D-l-sum-3}%
\end{align}
Now we use the $D\rightarrow4-D$ symmetry (\ref{D-4-D-symmetry-phi})
connecting the levels of $z(\beta,2+2k)$ and $z(\beta,2-2k)$. As was explained
in Sec. \ref{D-4-D-symmetry-section}, this symmetry is partial: when one
passes from $D=2-2k$ to $D=2+2k$ one loses the states described by the
solutions of the algebraic equation $R_{k-1}(E)=0$. Therefore%
\begin{equation}
z(\beta,2-2k)=z(\beta,2+2k)+\left[  \mathrm{Tr}\exp\left(  -\beta
H_{j=-(k-1)/2}\right)  \right]
\end{equation}
where $H_{j}$ is the matrix Hamiltonian (\ref{H-J-def}) corresponding to the
spin%
\begin{equation}
j=\frac{k-1}{2}\,. \label{j-via-n}%
\end{equation}
Now eq. (\ref{Z-D-l-sum-3}) takes the form%
\begin{align}
Z(\beta,-2M)  &  =\sum\limits_{k=1}^{M+1}\frac{(-1)^{k-M+1}2k\left[
(1+2M)!\right]  }{(M+1-k)!(1+M+k)!}\\
&  \times\mathrm{Tr}\exp\left(  -\beta H_{j=(k-1)/2}\right)  \,.
\end{align}
Let us replace the summation variable $k$ to $j$ (\ref{j-via-n})%
\begin{equation}
Z(\beta,-2M)=\sum\limits_{j=0}^{M/2}(-1)^{2j-M}\frac{2(2j+1)\left[
(1+2M)!\right]  }{(M-2j)!(2+M+2j)!}\,\mathrm{Tr}\exp\left(  -\beta
H_{j}\right)  \,. \label{Z-m2M-res}%
\end{equation}
Here the summation runs over integer and half-integer $j$.

\section{Fermion representation}

\label{Fermion-section}

\setcounter{equation}{0} 

\subsection{Derivation}

Let us start from the case of the quartic anharmonic oscillator
(\ref{H-quartic}). We can write the path integral for the partition function%

\begin{equation}
Z^{\mathrm{QO}}(\beta,D,g)=\int\limits_{q(0)=q(\beta)}Dx\exp\left\{  -\int
_{0}^{\beta}dt\left[  \frac{1}{2}\sum\limits_{i=1}^{D}\left(  \dot{x}_{i}%
^{2}+x_{i}^{2}\right)  +g\left(  \sum\limits_{i=1}^{D}x_{i}^{2}\right)
^{2}\right]  \right\}  \,. \label{path-int-1}%
\end{equation}
One can use the standard trick with the auxiliary field $\sigma$%
\begin{gather}
Z^{\mathrm{QO}}(\beta,D,g)=\int\limits_{q(0)=q(T)}Dx\int D\sigma\nonumber\\
\times\exp\left\{  -\int_{0}^{\beta}dt\left[  \frac{1}{2}\sum\limits_{i=1}%
^{D}x_{i}\left(  -\partial_{t}^{2}+1+i\sqrt{8g}\sigma\right)  x_{i}%
+\frac{\sigma^{2}}{2}\right]  \right\} \nonumber\\
=\int D\sigma\left[  \mathrm{Det}\left(  -\partial_{t}^{2}+1+i\sqrt{8g}%
\sigma\right)  \right]  ^{-D/2}\exp\left(  -\int_{0}^{\beta}dt\frac{\sigma
^{2}}{2}\right)  \,.
\end{gather}
For negative even $D$%
\begin{equation}
D=-2M \label{D-M}%
\end{equation}
we can use the fermion representation%
\begin{align}
&  \left[  \mathrm{Det}\left(  -\partial_{t}^{2}+1+i\sqrt{8g}\sigma\right)
\right]  ^{-D/2}\nonumber\\
&  =\int D\psi^{+}D\psi\exp\left\{  -\int_{0}^{\beta}dt\left[  \sum
\limits_{i=1}^{M}\psi_{i}^{+}\left(  -\partial_{t}^{2}+1+i\sqrt{8g}%
\sigma\right)  \psi_{i}\right]  \right\}  \,. \label{Det-int-psi}%
\end{align}
Integrating out the field $\sigma$, we obtain%
\begin{align}
&  Z^{\mathrm{QO}}(\beta,D,g)\overset{D\rightarrow-2M}{=}\int D\psi^{+}%
D\psi\nonumber\\
&  \times\exp\left\{  -\int_{0}^{\beta}dt\left[  \sum\limits_{i=1}^{M}\left[
\left(  \partial_{t}\psi_{i}^{+}\right)  \left(  \partial_{t}\psi_{i}\right)
+\psi_{i}^{+}\psi_{i}\right]  +4g\left(  \sum\limits_{i=1}^{M}\psi_{i}^{+}%
\psi_{i}\right)  ^{2}\right]  \right\}  \,.
\end{align}

The same trick can be repeated (at least formally) for an arbitrary polynomial
central potential $V(r^{2})$ if one uses more intricate integral
representations involving several auxiliary boson fields $\sigma$. In this way
one arrives at the fermion representation for partition function
(\ref{Z-beta-def}) with an arbitrary polynomial $V$:
\begin{align}
&  Z(\beta,D)\overset{D\rightarrow-2M}{=}\int D\psi^{+}D\psi\nonumber\\
&  \times\exp\left\{  -\int_{0}^{\beta}dt\left[  \sum\limits_{i=1}^{M}\left[
\left(  \partial_{t}\psi_{i}^{+}\right)  \left(  \partial_{t}\psi_{i}\right)
\right]  +V\left(  2\sum\limits_{i=1}^{M}\psi_{i}^{+}\psi_{i}\right)  \right]
\right\}  \,. \label{Z-fermion-general}%
\end{align}

Now we introduce the auxiliary fermion field $\chi_{i}$:%
\begin{align}
&  \int D\chi^{+}D\chi\exp\left\{  \int_{0}^{\beta}dt\sum\limits_{i=1}%
^{M}\left[  \left(  \partial_{t}\psi_{i}^{+}\right)  \chi_{i}-\chi_{i}%
^{+}\left(  \partial_{t}\psi_{i}\right)  -\chi_{i}^{+}\chi_{i}\right]
\right\} \nonumber\\
&  =\exp\left[  -\int_{0}^{\beta}dt\sum\limits_{i=1}^{M}\left(  \partial
_{t}\psi_{i}^{+}\right)  \left(  \partial_{t}\psi_{i}\right)  \right]  \,.
\end{align}
Then%
\begin{align}
&  Z(\beta,D)\overset{D\rightarrow-2M}{=}\int D\psi^{+}D\psi D\chi^{+}%
D\chi\nonumber\\
&  \times\exp\left\{  \int_{0}^{\beta}dt\left[  \sum\limits_{i=1}^{M}\left[
\left(  \partial_{t}\psi_{i}^{+}\right)  \chi_{i}-\chi_{i}^{+}\left(
\partial_{t}\psi_{i}\right)  -\chi_{i}^{+}\chi_{i}\right]  \right.  \right.
\nonumber\\
&  \left.  \left.  -V\left(  2\sum\limits_{i=1}^{M}\psi_{i}^{+}\psi
_{i}\right)  \right]  \right\}  \,.
\end{align}
Next we change the notation for the integration variables:%
\begin{align}
\psi_{i}  &  \rightarrow a_{i1}\,,\\
\psi_{i}^{+}  &  \rightarrow a_{i2}^{+}\,,\\
\chi_{i}  &  \rightarrow a_{i2}\,,\\
\chi_{i}^{+}  &  \rightarrow a_{i1}^{+}%
\end{align}
so that%
\begin{align}
&  Z(\beta,D)\overset{D\rightarrow-2M}{=}\int Da^{+}Da\nonumber\\
&  \times\exp\left\{  -\int_{0}^{\beta}dt\left[  \sum\limits_{i=1}^{M}%
\sum\limits_{\alpha,\beta=1}^{2}a_{i\alpha}^{+}\left(  \delta_{\alpha\beta
}\partial_{t}+\frac{1}{2}\left(  \sigma_{1}+i\sigma_{2}\right)  _{\alpha\beta
}\right)  a_{i\beta}\right.  \right. \nonumber\\
&  \left.  \left.  +V\left(  \sum\limits_{i=1}^{M}\sum\limits_{\alpha,\beta
=1}^{2}a_{i\alpha}^{+}\left(  \sigma_{1}-i\sigma_{2}\right)  _{\alpha\beta
}a_{i\beta}\right)  \right]  \right\}  \,. \label{Z-Grassmann}%
\end{align}
where $\sigma_{a}$ are standard Pauli matrices.

The integral on the RHS corresponds to the effective fermion Hamiltonian%
\begin{equation}
H_{F}=\sum\limits_{i=1}^{M}\sum\limits_{\alpha,\beta=1}^{2}\frac{1}%
{2}a_{i\alpha}^{+}\left(  \sigma_{1}+i\sigma_{2}\right)  _{\alpha\beta
}a_{i\beta}+V\left(  \sum\limits_{i=1}^{M}\sum\limits_{\alpha,\beta=1}%
^{2}a_{i\alpha}^{+}\left(  \sigma_{1}-i\sigma_{2}\right)  _{\alpha\beta
}a_{i\beta}\right)  \label{H-F-1}%
\end{equation}
with anticommutation relations%
\begin{align}
\left\{  a_{m\alpha},a_{n\beta}^{+}\right\}   &  =\delta_{mn}\delta
_{\alpha\beta}\,,\\
\left\{  a_{m\alpha},a_{n\beta}\right\}   &  =\left\{  a_{m\alpha}%
^{+},a_{n\beta}^{+}\right\}  =0\,.
\end{align}

Thus we have derived the formula%
\begin{equation}
Z(\beta,D)\overset{D\rightarrow-2M}{=}\mathrm{Tr}\,\left[  e^{-\beta H_{F}%
}P_{F}\right]  \label{Tr-continuation}%
\end{equation}
which should be understood in the sense of analytical continuation of the
bosonic partition function $\mathrm{Tr}\,e^{-\beta H}$ from positive integer
dimensions to negative even dimensions. On the RHS we have a trace in a
$2^{2M}$ dimensional vector space. $P_{F}$ is the operator counting the
fermion parity of the state:%
\begin{align}
P_{F}  &  =(-1)^{N_{F}+M}\,,\\
N_{F}  &  =\sum\limits_{i,\alpha}a_{i\alpha}^{+}a_{i\alpha}\,.
\end{align}
$P_{F}$ appears in eq. (\ref{Tr-continuation}) because the integral over
Grassmann variables (\ref{Z-Grassmann}) inherits periodic boundary conditions
of the original boson integral (\ref{path-int-1}).

Now we define%
\begin{equation}
\tilde{T}_{a}=\frac{1}{2}\sum\limits_{i=1}^{M}\sum\limits_{\alpha,\beta=1}%
^{2}a_{i\alpha}^{+}\left(  \sigma_{a}\right)  _{\alpha\beta}a_{i\beta}\,.
\end{equation}
Obviously%
\begin{equation}
\left[  \tilde{T}_{a},\tilde{T}_{b}\right]  =i\varepsilon_{abc}\tilde{T}_{c}%
\end{equation}
Introducing $sl(2)$ generators%
\begin{align}
\tilde{T}_{\pm}  &  =\tilde{T}_{1}\pm i\tilde{T}_{2}\,,\\
\tilde{T}_{0}  &  =\tilde{T}_{3}%
\end{align}
we can write Hamiltonian (\ref{H-F-1}) as%
\begin{equation}
H_{F}=\tilde{T}_{+}+V(2\tilde{T}_{-})\,. \label{H-F-def}%
\end{equation}
This effective Hamiltonian has a form similar to form (\ref{H-T-def}).

\subsection{Degeneracy factors}

Note that in the first derivation of the spin Hamiltonian $H_{j}$
(\ref{H-J-def}) via the $sl(2)$ expression $H_{T}$ (\ref{H-T-def}) we worked
with the ``effective dimension'' $\mathcal{D}$ given by (\ref{Delta-def})
keeping in mind that the spectrum depends on $D$ and $l$ only via
$\mathcal{D}$. On the contrary, in the derivation based on the fermion
Hamiltonian $H_{F}$ (\ref{H-F-def}) we can trace the contributions of levels
with different $l$. Strictly speaking, $H_{F}$ (\ref{H-F-def}) is not
completely equivalent to $H_{j}$ (\ref{H-J-def}) because $H_{F}$ acts in the
$2^{2M}$ dimensional space corresponding to the reducible representation of
$sl(2)$ (or $su(2)$ if one prefers a more familiar physical interpretation)%
\begin{equation}
\bigotimes\limits_{i=1}^{M}\left(  [0]%
{\textstyle\bigoplus}
[0]%
{\textstyle\bigoplus}
\left[  \frac{1}{2}\right]  \right)  \,. \label{tensor-product-1}%
\end{equation}
Here we use notation $[j]$ for irreducible representations corresponding to
spin $j$. Each factor $[0]%
{\textstyle\bigoplus}
[0]%
{\textstyle\bigoplus}
\left[  \frac{1}{2}\right]  $ in (\ref{tensor-product-1}) is associated with
the states in the $i$-th sector:%
\begin{align}
a_{i\beta}^{+}|0\rangle &  \rightarrow\left[  \frac{1}{2}\right]
\,,\label{one-particle-states}\\
|0\rangle &  \rightarrow\lbrack0]\,,\\
a_{i1}^{+}a_{i2}^{+}|0\rangle &  \rightarrow\lbrack0]\,.
\end{align}

Tensor product (\ref{tensor-product-1}) can be decomposed in irreducible
representations $[j]$%
\begin{equation}
\bigotimes\limits_{i=1}^{M}\left(  [0]%
{\textstyle\bigoplus}
[0]%
{\textstyle\bigoplus}
\left[  \frac{1}{2}\right]  \right)  =\bigoplus_{j=0}^{M/2}n(j,M)[j]
\label{N-product}%
\end{equation}
with degeneracies%
\begin{equation}
n(j,M)=\frac{2\left(  2j+1\right)  \Gamma(2M+2)}{\Gamma\left(  M-2j+1\right)
\Gamma(3+M+2j)} \label{n-j-M}%
\end{equation}
which can be computed using the recursion relation%
\begin{equation}
n(j,M+1)=2n(j,M)+n\left(  j-\frac{1}{2},M\right)  +n\left(  j+\frac{1}%
{2},M\right)  \,.
\end{equation}

Now we apply decomposition (\ref{N-product}) to the calculation of the
partition function (\ref{Tr-continuation}):%
\begin{equation}
Z(\beta,D)\overset{D\rightarrow-2M}{=}\sum\limits_{j=0}^{M/2}(-1)^{2j-M}%
n(j,M)\,\,\left[  \mathrm{Tr}\exp\left(  -\beta H_{j}\right)  \right]  \,.
\end{equation}
Inserting $n(j,M)$ from eq. (\ref{n-j-M}), we arrive at the result coinciding
with (\ref{Z-m2M-res}). Thus fermion representation (\ref{Z-fermion-general})
for the partition function leads to the same result as the direct calculation
of Sec. \ref{Analytical-PF-section}.

\subsection{Connection with the results of Dunne and Halliday}

If one uses representation%
\begin{align}
a_{i2}^{+}  &  =\theta_{2i-1}\,,\\
a_{i2}  &  =\frac{\partial}{\partial\theta_{2i-1}}\,,\\
a_{i1}  &  =\theta_{2i}\,,\\
a_{i1}^{+}  &  =\frac{\partial}{\partial\theta_{2i}}%
\end{align}
then Hamiltonian $H_{F}$ takes the form
\begin{equation}
H_{F}=\sum\limits_{i=1}^{M}\frac{\partial}{\partial\theta_{2i}}\frac{\partial
}{\partial\theta_{2i-1}}+V\left(  \sum\limits_{i=1}^{M}\theta_{2i-1}%
\theta_{2i}\right)
\end{equation}%
\begin{equation}
H_{F}=\sum\limits_{i=1}^{M}\sum\limits_{\alpha,\beta=1}^{2}\frac{1}%
{2}a_{i\alpha}^{+}\left(  \sigma_{1}+i\sigma_{2}\right)  _{\alpha\beta
}a_{i\beta}+V\left(  \sum\limits_{i=1}^{M}\sum\limits_{\alpha,\beta=1}%
^{2}a_{i\alpha}^{+}\left(  \sigma_{1}-i\sigma_{2}\right)  _{\alpha\beta
}a_{i\beta}\right)
\end{equation}
used by Dunne and Halliday \cite{DH-88}. The advantage of this form is the
possibility to work with the pseudo-boson variables%
\begin{align}
\xi_{i}  &  =\theta_{2i-1}\theta_{2i}\,,\\
\xi_{i}\xi_{k}  &  =-\xi_{k}\xi_{i}\,,\\
\left(  \xi_{i}\right)  ^{2}  &  =0
\end{align}
and to look for the eigenstates of $H_{F}$ in the form of polynomials
$\Psi(\{\xi_{i}\})$ in $\xi_{i}$.

In our representation these polynomial wave functions correspond to the states
of the form%
\begin{equation}
P(\left\{  a_{i2}^{+}a_{i1}\right\}  )|\Omega\rangle\, \label{P-Omega-states}%
\end{equation}
where $P$ are polynomials of $a_{i2}^{+}a_{i1}$, and $|\Omega\rangle$ is fixed
by the conditions%
\begin{equation}
a_{i1}^{+}|\Omega\rangle=a_{i2}|\Omega\rangle=0\quad(1\leq i\leq M)\,.
\end{equation}
This corresponds to working in the subspace of states obeying the constraint%
\begin{equation}
\sum_{\beta=1}^{2}a_{i\beta}^{+}a_{i\beta}\psi=\psi\quad(1\leq i\leq M)\,.
\end{equation}
In terms of decomposition (\ref{N-product}) this constraint selects states
(\ref{one-particle-states}) in each $i$-sector. Therefore states
(\ref{P-Omega-states}) belong to the subspace:%
\begin{equation}
\bigotimes\limits_{i=1}^{M}\left[  \frac{1}{2}\right]  =\bigoplus
_{j=\{M/2\}}^{M/2}\tilde{n}_{j}(j,M)[j]\,, \label{DH-decomposition}%
\end{equation}
where $\{x\}$ stands for the fractional part of $x$, and the degeneracy
factors%
\begin{equation}
\tilde{n}(j,M)=\frac{(-1)^{M+2j}+1}{2}\frac{(2j+1)M!}{\left(  \frac{M}%
{2}+j+1\right)  !\left(  \frac{M}{2}-j\right)  !}%
\end{equation}
follow from the recursion relation%

\begin{equation}
\tilde{n}(j,M+1)=\tilde{n}\left(  j-\frac{1}{2},M\right)  +\tilde{n}\left(
j+\frac{1}{2},M\right)  \,.
\end{equation}
Since Dunne and Halliday \cite{DH-88} (working with coupling constant
$\lambda=8g$) compute their characteristic polynomials $P_{M}(E,\lambda)$ in
the subspace associated with decomposition (\ref{DH-decomposition}), their
polynomials $P_{M}(E,\lambda)$ factorize into our polynomials $R_{M}%
(E,\lambda)$ (\ref{R-2j-def}):%
\begin{equation}
\left[  P_{M}(E,\lambda)\right]  _{\lambda=8g}=\mathrm{const}\,\prod
\limits_{j=\{M/2\}}^{M/2}\left[  R_{2j}(E,g)\right]  ^{\tilde{n}_{j}}\,.
\end{equation}
For example,%
\begin{align*}
P_{1}  &  =\mathrm{const}\,R_{1}\,,\\
P_{2}  &  =\mathrm{const}\,R_{0}R_{2}\,,\\
P_{3}  &  =\mathrm{const}\,R_{1}^{2}R_{3}\,,\\
P_{4}  &  =\mathrm{const}\,R_{0}^{2}R_{2}^{3}R_{4}\,.
\end{align*}
In this way the $sl(2)$ algebra explains the ``miracles'' observed in
\cite{DH-88}.

\section{Large orders of the perturbation theory}

\setcounter{equation}{0} 

\label{Higher-orders-section}

\subsection{Disappearance of the factorial divergence at negative even $D$}

\label{No-factorials-subsection}

In a general case the perturbative series for the energy of the ground state
of the $D$-dimensional quartic anharmonic oscillator (\ref{H-quartic})%

\begin{equation}
E(g,D)=\sum\limits_{k=0}^{\infty}E^{(k)}(D)g^{k} \label{E-g-expansion}%
\end{equation}
has factorially growing coefficients \cite{Zinn-Justin-81a}:%
\begin{align}
&  E^{(k)}(D)\overset{k\rightarrow\infty}{=}(-1)^{k+1}\Gamma\left(
k+\frac{D}{2}\right)  3^{k+\frac{D}{2}}\frac{2^{D/2}}{\pi\Gamma\left(
D/2\right)  }\nonumber\\
&  \times\left[  1-\frac{1}{6k}\left(  \frac{5}{3}+\frac{9}{2}D+\frac{7}%
{4}D^{2}\right)  +O(k^{-2})\right]  \,. \label{E-k-N-asymptotic}%
\end{align}
However, at negative even values of $D$ function $\left[  \Gamma\left(
D/2\right)  \right]  ^{-1}$ vanishes. Note that $\left[  \Gamma\left(
D/2\right)  \right]  ^{-1}$ appears on the RHS as a common factor so that at
negative even values of $D$ all terms of the systematic expansion in $1/k$
become zero. This hints that at $D=0,-2,-4$ the factorial growth of
perturbative coefficients disappears. The same happens in for the potential%
\begin{equation}
V(r^{2})=\frac{1}{2}r^{2}+gr^{2N}\,.
\end{equation}
In this case the factorial structure is different \cite{BLZ-77}%
\begin{align}
&  E^{(k)}(D)\overset{k\rightarrow\infty}{=}\left[  k(N-1)\right]  !\left[
-\frac{1}{\pi}\frac{(N-1)^{D/2}}{\Gamma(D/2)}\right]  \left[  \frac{\Gamma
\left(  \frac{2N}{N-1}\right)  }{\Gamma^{2}\left(  \frac{N}{N-1}\right)
}\right]  ^{D/2}\nonumber\\
&  \times k^{(D/2)-1}\left\{  \left[  -\frac{1}{2}\frac{\Gamma\left(
\frac{2N}{N-1}\right)  }{\Gamma^{2}\left(  \frac{N}{N-1}\right)  }\right]
^{N-1}\right\}  ^{k}\left[  1+O(k)\right]  \,.
\end{align}
but we still have the same factor of $\left[  \Gamma\left(  D/2\right)
\right]  ^{-1}$.

This factor of $\left[  \Gamma\left(  D/2\right)  \right]  ^{-1}$ also appears
in the asymptotic expression for $E^{(k)}(D)$ in the $O(D)$ symmetric model
studied in Ref. \cite{RS-95}. This model describes a particle on a
$D$-dimensional sphere with a potential breaking the $O(D+1)$ symmetry of the
sphere to $O(D)$.

One can easily trace the origin of this universal factor $\left[
\Gamma\left(  D/2\right)  \right]  ^{-1}$. In the semiclassical path integral
derivation of large-order asymptotic formulas, the common factor of $\left[
\Gamma\left(  D/2\right)  \right]  ^{-1}$ comes from the integration over
rotational collective coordinates of the classical solution breaking the
$O(D)$ symmetry of the problem to $O(D-1)$. As a result of this integration,
the final result is proportional to the surface area of $(D-1$)-dimensional
sphere in the $D$-dimensional space:%
\begin{equation}
S_{D}=\frac{2\,\pi^{D/2}}{\Gamma(D/2)}\,.
\end{equation}

Let us study the asymptotic behavior of $E^{(k)}(D)$ at negative even $D=-2M$.
We will consider the case when the potential has the form%
\begin{equation}
V(r)=\frac{1}{2}r^{2}+gU(r^{2})
\end{equation}
where $U(r^{2})$ is some polynomial. We know that at $D=-2M$ the energy of the
ground state (understood as usual in the sense of analytical continuation) is
given (at least for small $g$) by one of the roots of the algebraic equation
(\ref{R-M-g-spectrum-eq}). Since $R_{M}(E,g)$ is a polynomial in both $g$ and
$E$, the resulting dependence of $E(g,-2M)$ is regular for most values of $g$.
The radius of convergence of the perturbative expansion of $E(g,-2M)$ is
controlled by the singularity of $E(g,-2M)$ closest to $g=0$. These
singularities appear at values of $g$ corresponding to degenerate roots of
polynomials $R_{M}(E,g)$ which are described by equations%
\begin{equation}
R_{M}(E_{0},g_{0})=\left[  \frac{\partial}{\partial E}R_{M}(E,g_{0})\right]
_{E=E_{0}}=0\,. \label{R-M-degenerate-root}%
\end{equation}
Note that at $g=0$ polynomial $R_{M}(E,g)$ is given by expression
(\ref{det-H-H-HO}) which has no degenerate roots so that the point $g=0$ is
regular and we have a nonzero convergence radius. Equations
(\ref{R-M-degenerate-root}) may have several solutions. One has to choose the
solution corresponding to the singularity closest to the point $g=0$ (on the
main Riemann sheet). In the vicinity of the degenerate point $g_{0},E_{0}$ we
can use the Taylor expansion%
\begin{equation}
R_{M}(E,g)=\frac{1}{2}(E-E_{0})^{2}\left[  \frac{\partial^{2}}{\partial E^{2}%
}R_{M}(E,g_{0})\right]  _{E=E_{0}}+(g-g_{0})\left[  \frac{\partial}{\partial
g}R_{M}(E_{0},g)\right]  _{g=g_{0}}+\ldots
\end{equation}
Now equation (\ref{R-M-g-spectrum-eq}) leads to the following expression valid
in the vicinity of the singularity of $E(g)$ at $g=g_{0}$ is described by
equation%
\begin{equation}
E(g)\overset{g\rightarrow g_{0}}{=}E_{0}-\sqrt{c\left(  1-\frac{g}{g_{0}%
}\right)  } \label{E-root-singularity}%
\end{equation}
where%
\begin{equation}
c=-2g_{0}\left[  \frac{\partial}{\partial g}R_{M}(E_{0},g)\right]  _{g=g_{0}%
}\left\{  \left[  \frac{\partial^{2}}{\partial E^{2}}R_{M}(E,g_{0})\right]
_{E=E_{0}}\right\}  ^{-1}\,. \label{c-via-R-M}%
\end{equation}
Expanding%
\begin{equation}
\sqrt{1-\frac{g}{g_{0}}}=-\frac{1}{2\sqrt{\pi}}\sum\limits_{k=0}^{\infty
}\frac{\Gamma\left(  k-\frac{1}{2}\right)  }{\Gamma\left(  k+1\right)
}\left(  \frac{g}{g_{0}}\right)  ^{k}%
\end{equation}
and using the limit $k\rightarrow\infty$%
\begin{equation}
\frac{\Gamma\left(  k-\frac{1}{2}\right)  }{\Gamma\left(  k+1\right)
}\rightarrow k^{-3/2}\,,
\end{equation}
we find from eq. (\ref{E-root-singularity})%
\begin{equation}
E^{(k)}(D)\overset{k\rightarrow\infty}{\longrightarrow}\frac{1}{2}%
\sqrt{\frac{c(D)}{\pi}}k^{-3/2}\left[  g_{0}(D)\right]  ^{-k}\,\quad(D=-2M).
\label{E-k-D-asymptotic-D-special}%
\end{equation}

\subsection{Cases $D=0,-2,-4,-6$}

Now we want to concentrate on the case of the quartic anharmonic oscillator
(\ref{H-quartic}) and to consider cases $D=0,-2,-4,-6$.

The values $D=0$ and $D=-2$ are trivial because the corresponding polynomials
$\tilde{R}_{0}$ (\ref{R-0}) and $\tilde{R}_{1}$ (\ref{R-1}) do not depend on
$g$ so that the energy is given by%
\begin{align}
E(0)  &  =0\,,\\
E(-2)  &  =-1
\end{align}
and almost all perturbative coefficients $E^{(k)}$ vanish:%
\begin{align}
E^{(k)}(0)  &  =0\label{E-k-D0}\\
E^{(k)}(-2)  &  =-\delta_{k0} \label{E-k-Dm2}%
\end{align}

In the case $D=-4$ we insert polynomial $\tilde{R}_{2}(E,g)$ (\ref{R-2}) into
eq. (\ref{R-M-degenerate-root}) and find two solutions%

\begin{equation}
E_{0}(-4)=\mp\frac{2}{\sqrt{3}}\,,\quad g_{0}(-4)=\pm3^{-3/2}\,.
\label{E0-g0-m4-uncertain}%
\end{equation}
Taking into account that we are interested in the energy of the ground state
that corresponds to%

\begin{equation}
\left[  E(g,D)\right]  _{g=0}=\frac{D}{2}\,,
\end{equation}%
\begin{equation}
E(0,-4)=-2\,,
\end{equation}
one can check that the relevant singularity corresponds to the upper signs in
eq. (\ref{E0-g0-m4-uncertain})%
\begin{equation}
E_{0}(-4)=-\frac{2}{\sqrt{3}}\,,\quad g_{0}(-4)=3^{-3/2}\,.
\label{g0-solution}%
\end{equation}
Now we find from eq. (\ref{c-via-R-M})%
\begin{equation}
c(-4)=\frac{8}{9}\, \label{c-m4}%
\end{equation}
and insert this into eq. (\ref{E-k-D-asymptotic-D-special})%
\begin{equation}
E^{(k)}(-4)\overset{k\rightarrow\infty}{\longrightarrow}\frac{1}{3}%
\sqrt{\frac{2}{\pi}}\sum\limits_{k\gg1}^{\infty}k^{-3/2}3^{3k/2}g^{k}
\label{E-k-4-asymptotic}%
\end{equation}
for the coefficients $E^{(k)}(D)$ of expansion (\ref{E-series-0}) at $D=-4$.

Other cases of negative even $D$ can be analyzed in the same way. For example,
solving equation (\ref{R-M-degenerate-root}) for the polynomial $R_{3}(E,g)$
given by eq. (\ref{R-3}) we find for $D=-6$%
\begin{align}
g_{0}(-6)  &  =\frac{5\sqrt{13}-1}{36\sqrt{3\left(  5+2\sqrt{13}\right)  }%
}\,,\label{g0-minus-6}\\
E_{0}(-6)  &  =-\sqrt{\frac{1}{3}\left(  5+2\sqrt{13}\right)  }\,,\\
c(-6)  &  =\frac{2}{9}\left(  5-\frac{1}{\sqrt{13}}\right)  \label{c-m-6}%
\end{align}
which results in the asymptotic behavior%
\begin{equation}
E^{(k)}(-6)\overset{k\rightarrow\infty}{\longrightarrow}\frac{1}{6}%
\sqrt{\frac{2}{\pi}\left(  5-\frac{1}{\sqrt{13}}\right)  }k^{-3/2}\left[
g_{0}(-6)\right]  ^{-k}\,. \label{E-k-6-asymptotic}%
\end{equation}

\subsection{Roots of polynomials $E^{(k)}(D)$}

Coefficients $E^{(k)}(D)$ of the perturbative expansion (\ref{E-g-expansion})
are polynomials of degree $k+1$. Let us denote the $k+1$ roots of these
polynomials $\nu_{k,r}$:%
\begin{equation}
E^{(k)}(\nu_{k,r})=0\,\quad(1\leq r\leq k+1) \label{P-nu-root-eq}%
\end{equation}
According to eqs. (\ref{E-k-D0}), (\ref{E-k-Dm2}), at $k\geq1$ polynomials
$E^{(k)}(D)$ have a common factor of $D(D+2)$, i.e.%

\begin{equation}
E^{(k)}(D)=D(D+2)P_{k}(D)\,\, \label{E-via-P}%
\end{equation}
where $P_{k}(D)$ is a polynomial of degree $k-1$ with $k-1$ roots $\nu_{k,r}$.
This means that the roots $\nu_{k,r}$ include values $0$ and $-2$:%
\begin{equation}
\left\{  \nu_{k,r}\right\}  =\{0,-2,\ldots\}
\end{equation}

Note that the function $\left[  \Gamma\left(  D/2\right)  \right]  ^{-1}$ on
the RHS of eq. (\ref{E-k-N-asymptotic}) has zeros at%
\begin{equation}
D=0,-2,-4,-6,\ldots
\end{equation}
The roots of the asymptotic expression (\ref{E-k-N-asymptotic}) at $D=0,-2$
appear explicitly in the factorized expression on the RHS of eq.
(\ref{E-via-P}). The other roots should appear asymptotically at large $k$. In
other words, in the full set of roots $\nu_{k,r}$ there must be subsets
converging to values $D=-4,-6,\ldots$

Let us study the first subset converging to
\[
D=-2M\,\quad(M=-2,-3,\ldots)
\]
We label this subset of roots with the index $r=r(k,-2M)$:%
\begin{align}
E^{(k)}(\nu_{k,r(k,-2M)})  &  =0\,,\nonumber\\
\lim_{k\rightarrow\infty}\nu_{k,r(k,-2M)}  &  =-2M\,.
\end{align}

We want to derive a large-$k$ asymptotic formula for $2M+\nu_{k,r(k,-2M)}$. To
this aim we need an asymptotic formula for $E^{(k)}(D)$ in the double limit
$D\rightarrow-2M$ and $k\rightarrow\infty$. This asymptotic formula has the
form%
\begin{equation}
E^{(k)}(D)\overset{k\rightarrow\infty,\,D\rightarrow-2M}{\longrightarrow
}(-1)^{k+1}\Gamma\left(  k+\frac{D}{2}\right)  3^{k+\frac{D}{2}}\frac{2^{D/2}%
}{\pi\Gamma\left(  \frac{D}{2}\right)  }+E^{(k)}(-2M)\,.
\label{E-k-asymptotic-modified}%
\end{equation}
Note that we dropped the $1/k$ correction present in eq.
(\ref{E-k-N-asymptotic}) but added the extra term $E^{(k)}(-2M)$ which becomes
essential at $D\rightarrow-2M$ because of the suppressing factor $\left[
\Gamma\left(  \frac{D}{2}\right)  \right]  ^{-1}$ in the leading term. At
$D\rightarrow-2M$ we can simplify eq. (\ref{E-k-asymptotic-modified})%
\begin{equation}
E^{(k)}(D)\overset{k\rightarrow\infty,\,D\rightarrow-2M}{\longrightarrow
}(-1)^{k+1}k!k^{-M-1}3^{k-M}\frac{2^{-M}}{\pi\Gamma\left(  \frac{D}{2}\right)
}+E^{(k)}(-2M)\,. \label{E-k-asymptotic-modified-2}%
\end{equation}
Here%
\begin{equation}
\Gamma\left(  \frac{D}{2}\right)  \overset{D\rightarrow-2M}{\longrightarrow
}\frac{(-1)^{M}}{\frac{D}{2}+M}\frac{1}{M!}\,.
\end{equation}
Inserting this and eq. (\ref{E-k-D-asymptotic-D-special}) into eq.
(\ref{E-k-asymptotic-modified-2}), we obtain%
\begin{align}
&  E^{(k)}(D)\overset{k\rightarrow\infty,\,D\rightarrow-2M}{\longrightarrow
}(-1)^{k+1}k!k^{-M-1}3^{k-M}\frac{(-1)^{M}2^{-M}M!}{\pi}\left(  \frac{D}%
{2}+M\right) \nonumber\\
&  +\frac{1}{2}\sqrt{\frac{c(-2M)}{\pi}}k^{-3/2}\left[  g_{0}(-2M)\right]
^{-k}\,.
\end{align}
Using this asymptotic formula, we solve equation%
\begin{equation}
E^{(k)}(\nu_{k,r(k,-2M)})\overset{\nu_{k,r(k,-2M)}\rightarrow-2M}{=}0\,
\end{equation}
and find%
\begin{equation}
\nu_{k,r(k,-2M)}+2M\overset{k\rightarrow\infty}{\longrightarrow}\frac{\left(
-6\right)  ^{M}}{M!}\sqrt{\pi c(-2M)}\frac{(-1)^{k}}{k!}k^{M-(1/2)}\left[
3g_{0}(-2M)\right]  ^{-k}\,.
\end{equation}
Using above expressions (\ref{E0-g0-m4-uncertain}), (\ref{c-m4}),
(\ref{g0-minus-6}), (\ref{c-m-6}), for $c(-2M)$ and $g_{0}(-2M)$ at $M=2,3$,
we obtain%
\begin{equation}
\nu_{k,r(k,-4)}+4\overset{k\rightarrow\infty}{\longrightarrow}12\sqrt{2\pi
}\frac{(-1)^{k}}{k!}k^{3/2}3^{k/2}\,\,, \label{nu-k-asymptotic-4}%
\end{equation}%
\begin{equation}
\nu_{k,r(k,-6)}+6\overset{k\rightarrow\infty}{\longrightarrow}12\sqrt
{2\pi\left(  5-\frac{1}{\sqrt{13}}\right)  }\frac{(-1)^{k+1}}{k!}%
k^{5/2}\left[  \frac{5\sqrt{13}-1}{12\sqrt{3\left(  5+2\sqrt{13}\right)  }%
}\right]  ^{-k}\,. \label{nu-k-asymptotic-6}%
\end{equation}

Asymptotic formulas (\ref{nu-k-asymptotic-4}), (\ref{nu-k-asymptotic-6}) agree
with results of the direct numerical calculation using methods of Refs.
\cite{BW-69,BW-73,SZ-79,Zinn-Justin-81a}. The roots approaching $-4$ appear
rather early. They are listed in Table \ref{table-1}. For small odd values
$k=5,7,9$ these roots have small imaginary parts. Starting from $k=10$, the
roots $\nu_{k,r(k,4)}$ close to $-4$ become real and exhibit a very fast
factorial convergence to $-4$ which is described by the asymptotic formula
(\ref{nu-k-asymptotic-4}). Already at $k=11$ this formula works with accuracy
better than $7\%$.

\begin{table}[ptb]%
\begin{tabular}
[c]{||l|l||l|l||}\hline\hline
$k$ & $\nu_{k,r(k,-4)} +4 $ & $k$ & $\nu_{k,r(k,-4)} +4 $\\\hline
$5$ & $-3.22834\pm i0.426293 $ & $33$ & $-4.79523448 \times10^{-26} $\\\hline
$6$ & $-3.44545 $ & $34$ & $+2.55611474 \times10^{-27} $\\\hline
$7$ & $-3.63083\pm i0.34226 $ & $35$ & $-1.32186040 \times10^{-28} $\\\hline
$8$ & $-3.76443 $ & $36$ & $+6.63762841 \times10^{-30} $\\\hline
$9$ & $-3.9583\pm i0.226557 $ & $37$ & $-3.23912410 \times10^{-31} $\\\hline
$10$ & $+0.04231592827 $ & $38$ & $+1.53735607 \times10^{-32} $\\\hline
$11$ & $-0.01231265412 $ & $39$ & $-7.10198670 \times10^{-34} $\\\hline
$12$ & $+0.00178433080 $ & $40$ & $+3.19560318 \times10^{-35} $\\\hline
$13$ & $-0.00027422590 $ & $41$ & $-1.40147930 \times10^{-36} $\\\hline
$14$ & $+0.00003787462 $ & $42$ & $+5.99459657 \times10^{-38} $\\\hline
$15$ & $-4.86252995 \times10^{-6} $ & $43$ & $-2.50229008 \times10^{-39}
$\\\hline
$16$ & $+5.80950053 \times10^{-7} $ & $44$ & $+1.01993415 \times10^{-40}
$\\\hline
$17$ & $-6.49387664 \times10^{-8} $ & $45$ & $-4.06166432 \times10^{-42}
$\\\hline
$18$ & $+6.81906230 \times10^{-9} $ & $46$ & $+1.58111340 \times10^{-43}
$\\\hline
$19$ & $-6.75145346 \times10^{-10}$ & $47$ & $-6.01961315 \times10^{-45}
$\\\hline
$20$ & $+6.32321355 \times10^{-11}$ & $48$ & $+2.24249068 \times10^{-46}
$\\\hline
$21$ & $-5.61842521 \times10^{-12}$ & $49$ & $-8.17805838 \times10^{-48}
$\\\hline
$22$ & $+4.74864227 \times10^{-13}$ & $50$ & $+2.92092144 \times10^{-49}
$\\\hline
$23$ & $-3.82680164 \times10^{-14}$ & $51$ & $-1.02217301 \times10^{-50}
$\\\hline
$24$ & $+2.94682238 \times10^{-15}$ & $52$ & $+3.50623616 \times10^{-52}
$\\\hline
$25$ & $-2.17261017 \times10^{-16}$ & $53$ & $-1.17934411 \times10^{-53}
$\\\hline
$26$ & $+1.53640755 \times10^{-17}$ & $54$ & $+3.89122545 \times10^{-55}
$\\\hline
$27$ & $-1.04388321 \times10^{-18}$ & $55$ & $-1.25990087 \times10^{-56}
$\\\hline
$28$ & $+6.82474923 \times10^{-20}$ & $56$ & $+4.00444478 \times10^{-58}
$\\\hline
$29$ & $-4.29961693 \times10^{-21}$ & $57$ & $-1.24982870 \times10^{-59}
$\\\hline
$30$ & $+2.61370107 \times10^{-22}$ & $58$ & $+3.83179265 \times10^{-61}
$\\\hline
$31$ & $-1.53497157 \times10^{-23}$ & $59$ & $-1.15433788 \times10^{-62}
$\\\hline
$32$ & $+8.71891707 \times10^{-25}$ & $60$ & $+3.41802128 \times10^{-64}
$\\\hline\hline
\end{tabular}
\caption{Roots $\nu_{k,r(k,-4)}$ of polynomials $P_{D}(N)$ approaching the
value $-4$.}%
\label{table-1}%
\end{table}

At large values of $k$ there appear roots converging to other integer values
$D=-6,-8,\ldots$ The roots approaching the value $-6$ are listed in Table
\ref{table-2}. At large $k$ they are well described by asymptotic formula
(\ref{nu-k-asymptotic-6}).

\begin{table}[ptb]%
\begin{tabular}
[c]{||l|l||l|l||}\hline\hline
$k$ & $\nu_{k,r(k,-6)} +6 $ & $k$ & $\nu_{k,r(k,-6)} +6 $\\\hline
$21$ & $+0.03432898 $ & $41$ & $+1.44684 \times10^{-18} $\\\hline
$22$ & $-0.011016 $ & $42$ & $-1.55969 \times10^{-19} $\\\hline
$23$ & $+0.0020276 $ & $43$ & $+1.63996 \times10^{-20} $\\\hline
$24$ & $-0.00040899 $ & $44$ & $-1.68294 \times10^{-21} $\\\hline
$25$ & $+0.000076759 $ & $45$ & $+1.68654 \times10^{-22} $\\\hline
$26$ & $-0.000013876 $ & $46$ & $-1.65141 \times10^{-23} $\\\hline
$27$ & $+2.40442 \times10^{-6} $ & $47$ & $+1.58077 \times10^{-24} $\\\hline
$28$ & $-4.00519 \times10^{-7} $ & $48$ & $-1.47999 \times10^{-25} $\\\hline
$29$ & $+6.42173 \times10^{-8} $ & $49$ & $+1.35591 \times10^{-26} $\\\hline
$30$ & $-9.92472 \times10^{-9} $ & $50$ & $-1.21614 \times10^{-27} $\\\hline
$31$ & $+1.48040 \times10^{-9} $ & $51$ & $+1.06834 \times10^{-28} $\\\hline
$32$ & $-2.13385 \times10^{-10} $ & $52$ & $-9.19592 \times10^{-30} $\\\hline
$33$ & $+2.97550 \times10^{-11} $ & $53$ & $+7.75910 \times10^{-31} $\\\hline
$34$ & $-4.01814 \times10^{-12} $ & $54$ & $-6.41992 \times10^{-32} $\\\hline
$35$ & $+5.26006 \times10^{-13} $ & $55$ & $+5.21090 \times10^{-33} $\\\hline
$36$ & $-6.68131 \times10^{-14} $ & $56$ & $-4.15065 \times10^{-34} $\\\hline
$37$ & $+8.24173 \times10^{-15} $ & $57$ & $+3.24558 \times10^{-35} $\\\hline
$38$ & $-9.88147 \times10^{-16} $ & $58$ & $-2.49222 \times10^{-36} $\\\hline
$39$ & $+1.15242 \times10^{-16} $ & $59$ & $+1.87991 \times10^{-37} $\\\hline
$40$ & $-1.30830 \times10^{-17} $ & $60$ & $-1.39342 \times10^{-38}
$\\\hline\hline
\end{tabular}
\caption{Roots $\nu_{k,r(k,-6)}$ of polynomials $P_{k}(D)$ approaching the
value $-6$.}%
\label{table-2}%
\end{table}

\subsection{Distribution of roots in the complex plane}

\begin{figure}
[ptb]
\begin{center}
\includegraphics[
height=3.4221in,
width=3.3486in
]%
{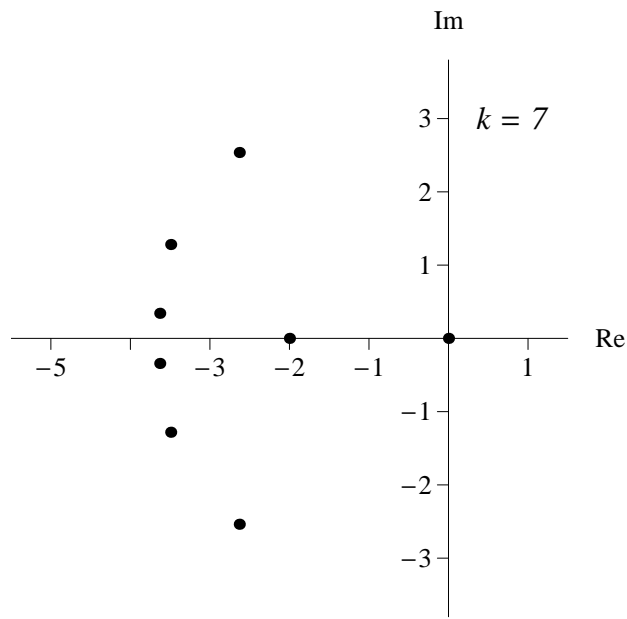}%
\caption{8 roots $\nu_{8,r}$ of the polynomial $E^{(7)}(D)$. Two complex
conjugate roots $-3.63\pm i0.34$ are in the vicinity of the negative even
value $D=-4$.}%
\label{roots-7}%
\end{center}
\end{figure}

\begin{figure}
[ptb]
\begin{center}
\includegraphics[
height=3.2785in,
width=3.4221in
]%
{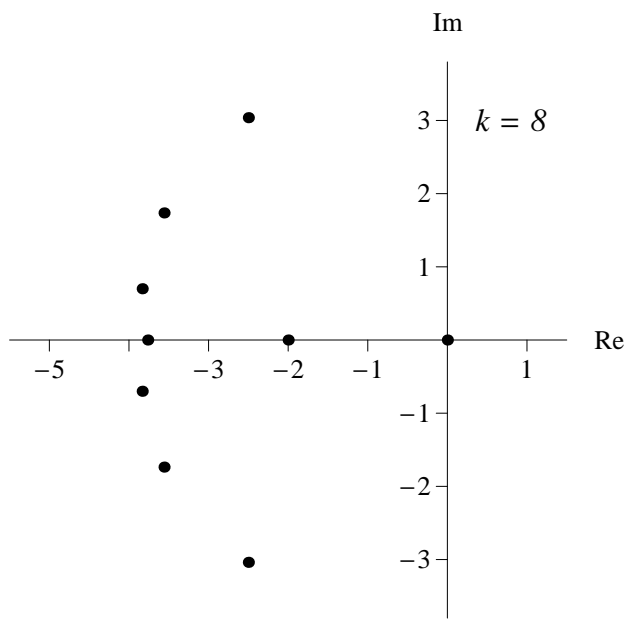}%
\caption{9 roots $\nu_{9,r}$ of the polynomial $E^{(8)}(D)$. One root at
$-3.76443$ is close to the negative even value $-4$.}%
\label{roots-8}%
\end{center}
\end{figure}

\begin{figure}
[ptb]
\begin{center}
\includegraphics[
height=3.2785in,
width=3.3944in
]%
{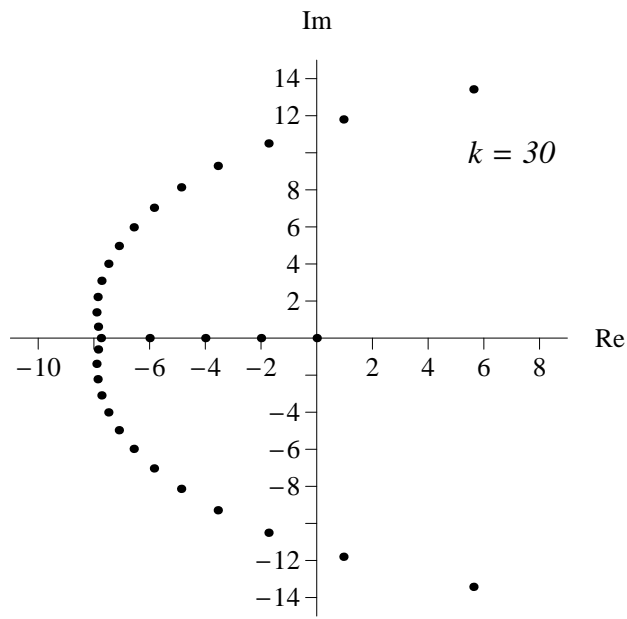}%
\caption{31 roots $\nu_{30,r}$ of the polynomial $E^{(30)}(D)$. One can see a
discrete set of roots close to even negative values up to $D=-8$. Most of the
roots belong to the quasicontinuous set formed in the complex plane.}%
\label{roots-30}%
\end{center}
\end{figure}

\begin{figure}
[ptb]
\begin{center}
\includegraphics[
height=3.2379in,
width=3.3797in
]%
{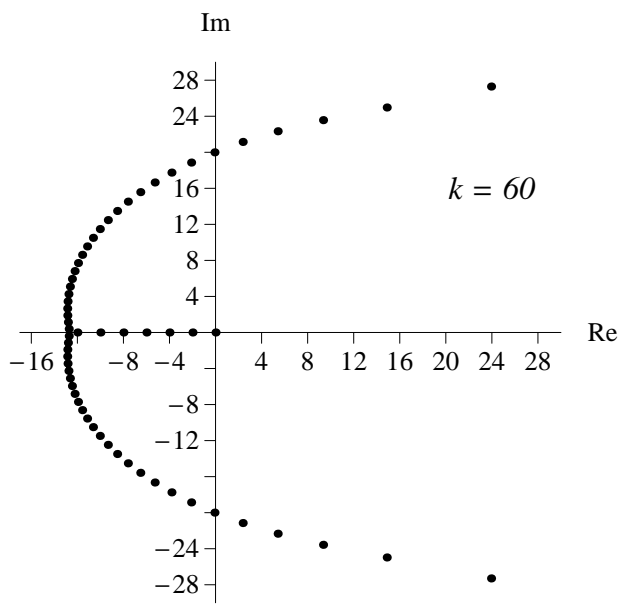}%
\caption{61 roots $\nu_{60,r}$ of the polynomial $E^{(60)}(D)$. One can see a
discrete set of roots close to even negative values up to $D=-12$. There is
also a quasicontinuous set formed in the complex plane.}%
\label{roots-60}%
\end{center}
\end{figure}

One should keep in mind that apart from the stable roots at $D=0,-2$ and the
roots asymptotically approaching points $D=-4,-6,\ldots$, there are many other
roots $\nu_{k,r}$. Most of them are complex. In order to get an impression
about the general distribution of roots in the complex plane, we show them for
the cases $k=7$ (Fig. \ref{roots-7}), $k=8$ (Fig. \ref{roots-8}), and $k=30$
(Fig. \ref{roots-30}) and $k=60$ (Fig. \ref{roots-60}). From the plots for
$k=30$ and $k=60$, it is clearly seen that at large $k$ there appears a
certain regular structure in the distribution of complex roots. It is an
interesting problem to give a complete asymptotic description of roots in the
complex plane. Here we make only preliminary comments about the distribution
of roots.

It is convenient to work with polynomials $P_{k}(D)$ (\ref{E-via-P}). Using
eq. (\ref{E-k-asymptotic-0}), we find:%
\begin{equation}
P_{k}(D)\overset{k\rightarrow\infty}{\longrightarrow}(-1)^{k+1}\Gamma\left(
k+\frac{D}{2}\right)  3^{k+\frac{D}{2}}\frac{2^{(D/2)-2}}{\pi\Gamma\left(
D/2+2\right)  }\,. \label{P-asymptotic}%
\end{equation}
The polynomial $P_{k}(D)$ has roots $\tilde{\nu}_{k,r}$ which coincide with
the roots of $E^{(k)}(D)$ up to the two stable roots $0$, $-2$:%
\[
\{\nu_{k,1},\nu_{k,2},\ldots\nu_{k,k+1}\}=\{\tilde{\nu}_{k,1},\tilde{\nu
}_{k,2},\ldots\tilde{\nu}_{k,k-1},0,-2\}\,.
\]
We can represent $P_{k}(D)$ in the form%
\begin{equation}
P_{k}(D)=\beta_{k}\prod\limits_{r=1}^{k-1}\left(  D-\tilde{\nu}_{k,r}\right)
\,. \label{P-root-product}%
\end{equation}
The coefficient $\beta_{k}$ was computed in Ref. \cite{DP-79}:%
\begin{equation}
\beta_{k}=(-1)^{k+1}2^{k-2}\frac{\Gamma(\frac{3k-1}{2})}{(k+1)!\Gamma\left(
\frac{k+1}{2}\right)  }\,. \label{beta-k-DP}%
\end{equation}
We find from eqs. (\ref{P-asymptotic}), (\ref{P-root-product})%

\begin{equation}
\frac{P_{k}(D)}{P_{k}(0)}=\prod\limits_{r=1}^{k-1}\left(  1-\frac{D}%
{\tilde{\nu}_{k,r}}\right)  \overset{k\rightarrow\infty}{\longrightarrow
}\frac{\left(  6k\right)  ^{D/2}}{\Gamma\left(  D/2+2\right)  }\,.
\end{equation}
Differentiating the logarithm of this expression with respect to $D$, we
obtain%
\begin{equation}
\sum\limits_{r=1}^{k-1}\frac{1}{D-\tilde{\nu}_{k,r}}\overset{k\rightarrow
\infty}{\longrightarrow}\frac{1}{2}\left[  \ln\left(  6k\right)  -\psi\left(
\frac{D}{2}+2\right)  \right]
\end{equation}
where $\psi(z)$ is the logarithmic derivative of the $\Gamma$ function%
\begin{equation}
\psi(z)=\frac{\Gamma^{\prime}(z)}{\Gamma(z)}=-\gamma-\sum\limits_{k=0}%
^{\infty}\left(  \frac{1}{z+k}-\frac{1}{k+1}\right)
\end{equation}
and $\gamma$ is Euler's constant. Thus%
\begin{equation}
\prod\limits_{r=1}^{k-1}\frac{1}{D-\tilde{\nu}_{k,r}}\overset{k\rightarrow
\infty}{\longrightarrow}\frac{1}{2}\ln\left(  6k\right)  +\frac{1}{2}\left[
\gamma+\sum\limits_{k=0}^{\infty}\left(  \frac{2}{D+4+2k}-\frac{1}%
{k+1}\right)  \right]  \,.
\end{equation}
This relation requires the appearance of roots $\nu_{k,r}$ converging to
$-4,-6,$\ldots\ but does not forbid the existence of additional complex roots
going to infinity or making a quasicontinuous distribution in the limit
$k\rightarrow\infty$. Numerical calculations show a large amount of complex
roots at large $k$ (see Figs. \ref{roots-30} and \ref{roots-60}).

In the limit of large $k$, one can compute the product of all roots $\nu
_{k,r}$ of the polynomials $P_{k}(D)$ setting $D=0$ in eqs.
(\ref{P-asymptotic}), (\ref{P-root-product}) and using expression
(\ref{beta-k-DP}) for $\beta_{k}$:%
\begin{equation}
\prod\limits_{r=1}^{k-1}\nu_{k,r}\overset{k\rightarrow\infty}{\longrightarrow
}\frac{1}{\beta_{k}}\frac{3^{k}}{4\pi}\Gamma\left(  k\right)  \overset
{k\rightarrow\infty}{\longrightarrow}(-1)^{k+1}3k^{2}\left(  \frac{k}%
{e\sqrt{3}}\right)  ^{k}\,.
\end{equation}
This gives us an asymptotic estimate for the geometric average of all nonzero
$k$ roots $\nu_{k,r}$ of $E_{k}(D)$ (including the root $\nu_{k,r}=-2$)%
\begin{equation}
\left\langle |\nu_{k,r}|\right\rangle _{\nu_{k,r}\neq0}\equiv\left[  2\left(
\prod\limits_{r=1}^{k-1}|\tilde{\nu}_{k,r}|\right)  \right]  ^{1/k}%
\overset{k\rightarrow\infty}{\longrightarrow}\frac{k}{e\sqrt{3}}\left(
6k^{2}\right)  ^{1/k}\,.
\end{equation}
The growth of this quantity with $k$ shows that most of the roots go to infinity.%

\section{Conclusions}

\setcounter{equation}{0} 

Thus we have established a close connection between

1) exactly solvable features of the anharmonic oscillator in negative even dimensions,

2) disappearance of the factorial growth of perturbative coefficients in
negative even dimensions,

3)\ fast inverse factorial convergence of roots of polynomials $E^{(k)}(D)$ to
the negative even points.

Although the content of this paper was restricted to quantum mechanics, the
main motivation comes from quantum field theory. In contrast to quantum
mechanics where the large-order behavior of the perturbation theory can be
easily tested using direct numerical calculations in cases when the analytical
methods fail or raise doubts, in quantum field theory the power of both
analytical and numerical tools is rather limited. The slow $O(k^{-1})$
convergence of perturbative coefficients to the asymptotic form [see e.g. eq.
(\ref{E-k-N-asymptotic})] is rather disturbing in quantum mechanics but one
still can reach the asymptotic regime in large orders. In quantum field theory
calculations rarely go beyond four or five loops. In this situation the
construction of quantities whose perturbative expansion has a fast factorial
convergence to the asymptotic form is very important.

Polynomials $E^{(k)}(D)$ and their roots have analogs in quantum field theory.
In many field theoretical models we have dependence on various parameters
(e.g. number of colors, flavors etc.) and Feynman diagrams in any order lead
to a polynomial (or fractional polynomial) dependence on these parameters. The
control of this polynomial dependence is usually trivial compared to the hard
work needed for the calculation of loop integrals. Therefore available
multiloop results in quantum field theory provide many opportunities for
testing the roots of these polynomials. There is some evidence that the
inverse factorial convergence of roots found in the case of the anharmonic
oscillator has analogs in some field theoretical models, e.g. in the
$N$-component $O(N)$ symmetric model \cite{PP-08}. An indirect theoretical
argument comes from the presence of an inverse $\Gamma$ function in field
theoretical asymptotic expressions. This inverse $\Gamma$ function may
generate zeros similar to the zeros at $D=0,-2,-4\ldots$ coming from the
factor $\left[  \Gamma(D)\right]  ^{-1}$ in eq. (\ref{E-k-asymptotic-0}) in
the case of the anharmonic oscillator. Since the convergence is very fast, it
may happen that one can detect it (or its signals) in available multi-loop
perturbative expressions.

\textbf{Acknowledgments.} I appreciate discussions with G.V. Dunne, L.N.
Lipatov, V.Yu. Petrov and A. Turbiner.

\end{document}